\documentclass[12pt]{article}

\usepackage{amsmath,amsfonts,amssymb,amscd}
\usepackage{color}
\usepackage[utf8]{inputenc}

\usepackage[small,nohug,heads=vee]{diagrams}
\diagramstyle[labelstyle=\scriptstyle]

\makeatletter
\newcommand\xleftrightarrow[2][]{%
  \ext@arrow 9999{\longleftrightarrowfill@}{#1}{#2}}
\newcommand\longleftrightarrowfill@{%
  \arrowfill@\leftarrow\relbar\rightarrow}
\makeatother

\newlength{\minitwocolumn}
\newcommand{\sbv}[2]{{\{{{#1},{#2}}\}}}

\newcommand{\cp}{{\check{p}}}
\newcommand{\cq}{{\check{q}}}
\newcommand{\cx}{{\check{x}}}
\newcommand{\cxi}{{\check{\xi}}}

\newcommand{\tB}{{\tilde{B}}}
\newcommand{\tbeta}{{\tilde{\beta}}}

\newcommand\proj{{pr}}
\newcommand{\hj}{{\widehat{j}}}
\newcommand{\hcM}{{\widehat{\mathcal{M}}}}
\newcommand{\hM}{{\widehat{M}}}

\newcommand{\cM}{{\mathcal{M}}}
\newcommand{\tM}{{\tilde{M}}}
\newcommand{\tq}{{\tilde{q}}}
\newcommand{\tp}{{\tilde{p}}}
\newcommand{\txi}{{\tilde{\xi}}}
\newcommand{\tx}{{\tilde{x}}}

\newcommand{\tj}{{\tilde{j}}}

\newcommand{\tpartial}{{\tilde{\partial}}}
\newcommand{\td}{{\tilde{d}}}
\newcommand{\te}{{\tilde{e}}}
\newcommand{\tu}{{\tilde{u}}}
\newcommand{\tpar}{{\tilde{\partial}}}
\newcommand{\Sp}{{\;\,}}

\newcommand{\wcM}{{\widehat{\cM}}}

\def\bR{{\mathbb{R}}}

\begin{document}

\baselineskip 0.7cm
\begin{titlepage}

\begin{flushright}
\null \hfill Preprint TU-1036\\[3em]
\end{flushright}

\begin{center}
{\Large \bf Unified picture of non-geometric fluxes and T-duality in double field theory 
via graded symplectic manifolds}
\vskip 1.2cm
Marc Andre Heller${}^{a,}$\footnote{E-mail:\
heller@tuhep.phys.tohoku.ac.jp}, Noriaki Ikeda${}^{b,}$\footnote{E-mail:\
nikeda@se.ritsumei.ac.jp
}
 ~and Satoshi Watamura${}^{a,}$\footnote{E-mail:\ watamura@tuhep.phys.tohoku.ac.jp}

\vskip 0.4cm
{

\it
${}^a$
Particle Theory and Cosmology Group, \\
Department of Physics, Graduate School of Science, \\
Tohoku University \\
Aoba-ku, Sendai 980-8578, Japan \\

\vskip 0.4cm
${}^b$
Department of Mathematical Sciences,
Ritsumeikan University \\
Kusatsu, Shiga 525-8577, Japan \\

}
\vskip 0.4cm

\vskip 1.5cm

\begin{abstract}
We give a systematic derivation of the local expressions of the NS $H$-flux, geometric $F$- as well as non-geometric $Q$- and $R$-fluxes in terms of bivector
$\beta$- and two-form $B$-potentials including vielbeins. They are obtained using a supergeometric method on QP-manifolds by twist of the standard Courant algebroid on the generalized tangent space without flux. Bianchi identities of the fluxes are easily deduced. We extend the discussion to the case of the double space and present a formulation of T-duality in terms of canonical transformations between graded symplectic manifolds. Finally, the construction is compared to the formerly introduced Poisson Courant algebroid, a Courant algebroid on a Poisson manifold, as a model for $R$-flux.
\end{abstract}
\end{center}
\end{titlepage}

\setcounter{page}{2}

\rm

\newpage

\tableofcontents

\newpage

\section{Introduction}
\noindent

There are two dualities interrelating the various ten-dimensional superstring theories: T-duality and S-duality. Whereas S-duality relates strong and weak coupling regimes, T-duality exchanges winding and momentum modes of closed strings wrapping compact cycles and is a map between different string backgrounds. It is a target space symmetry.

Additionally, fluxes wrapping the internal cycles of compactified string theories play an important role when considering T-duality. There is the NS-NS two-form $B$-field, to which the string couples, and its three-form field strength, the so-called $H$-flux. Furthermore, the $f$-flux is the torsion-less part of the projected spin-connection and therefore is closely related to the geometry of the compactified space itself.

In the case, where the compactified space exhibits a Killing isometry, T-duality in this direction is possible and mixes $B$-field and metric components. The equations that express the new metric and $B$-field in terms of the old ones are the so-called Buscher rules \cite{Buscher1,Buscher2}.

However, if one considers successive T-duality transformations of a three-torus with $H$-flux background, so-called non-geometric backgrounds with associated non-geometric fluxes $Q$ and $R$ appear \cite{Shelton:2005cf,Wecht:2007ngf}. The $Q$-flux signalizes a globally non-geometric background with monodromy, which has to be patched by T-duality transformation. The $R$-flux signalizes an even locally non-geometric background, where standard manifold descriptions fail.

Double field theory \cite{Hull:2009mi,Aldazabal:2011nj} approaches this problem by the introduction of winding coordinates, which are dual to the standard ones, and formulating T-duality on toroidal backgrounds as an $O(D,D)$-transformation on this doubled set of coordinates. In this formulation, even T-duality in non-isometry directions is possible and the non-geometric $Q$- and $R$-flux can be interpreted naturally \cite{Hull:2004in,Andriot:2012an}. The potential of $R$-flux is conjectured to be given by a bivector field $\beta$. A supergravity formulation making use of the $\beta$-potential can be found in \cite{Andriot:2013xca}.

Since T-duality mixes metric and $B$-field, both structures can be combined in an $O(D,D)$-tensor, the so-called generalized metric. It turns out that the associated backgrounds with $H$-flux can be naturally described using the Courant algebroid on the generalized tangent bundle $TM\oplus T^*M$. The underlying structure is given by generalized geometry \cite{Gualtieri:2003dx,Cavalcanti:2011wu}. The geometric subgroup of T-duality transformations, spanned by diffeomorphisms and $B$-transformations, leave the inner product on the Courant algebroid invariant. However, under $B$-transformation the Courant bracket gets twisted by a term proportional to $dB=H$. An analysis of non-geometric backgrounds and their relation to generalized geometry is conducted in \cite{Grana:2008yw}.

A different model realizing $R$-flux was given in \cite{Asakawa:2014kua} using a Poisson tensor. This model is called Poisson Courant algebroid and the Poisson tensor is used to define the Courant algebroid on the generalized tangent bundle, where the roles of the tangent and cotangent spaces are exchanged. The associated structure is called Poisson-generalized geometry. Along this line, the authors of \cite{Bessho:2015tkk} reformulated the Poisson Courant algebroid using graded symplectic manifolds and elucidated its relation to $H$-flux backgrounds as well as to double field theory.

From the topological perspective, T-duality on toroidal backgrounds with $H$-flux was analyzed in \cite{Bouwknegt:2003vb,Bouwknegt:2004tr}. An analogous examination in the setting of Poisson-generalized geometry including $Q$-flux was carried out in \cite{Asakawa:2015aia}. From the perspective of graded symplectic manifolds, the structure of double field theory was analyzed in \cite{Deser:2014mxa,DeserSaemann:2016}. $\alpha$'-corrections to the C-bracket in double field theory from the viewpoint of deformations of graded manifolds were discussed in \cite{Deser:2014ap}. Relations of non-geometric fluxes to non-associative geometry were discussed in \cite{Blumenhagen:2011ph,Blumenhagen:2010hj}. From the perspective of membrane sigma-models, a string in an $R$-flux background was proposed to propagate in the cotangent bundle $T^*M$, which upon quantization develops a non-associative Moyal-Weyl-type star product \cite{Mylonas:2012pg}. The non-associativity is conjectured to be governed by the $R$-flux.

In this paper, we show how the local expression for the geometric $H$- and $F$- as well as the non-geometric $Q$- and $R$-fluxes can be introduced naturally from twists of the Courant algebroid on $TM\oplus T^*M$ without flux. For this, we make use of QP-manifolds of degree two, which naturally are linked to general Courant algebroids \cite{Roytenberg99}. The twists correspond to $B$-transformations, $\beta$-transformations and diffeomorphisms and therefore span the full $O(D,D)$ T-duality group. In order to introduce twists representing the diffeomorphisms, a frame bundle is defined. These twists then naturally introduce vielbeins into the description. The operations of the resulting twisted Courant algebroid are induced by derived brackets and the graded Poisson bracket of the QP-manifold. The classical master equation on the QP-manifold is shown to lead to the Bianchi identities among the fluxes. 

We then derive the local expressions of all the fluxes in the setting of double field theory by twist of the Hamiltonian analyzed in \cite{Deser:2014mxa}. The associated structure is called pre-QP-manifold in \cite{DeserSaemann:2016}. By construction, the resulting algebraic structure can be projected by solving the section condition to either give a twisted Courant algebroid in the supergravity frame, or a twisted Courant algebroid that lives entirely in the winding frame. Of course, mixed solutions of the sections condition give rise to different twisted Courant algebroids. Again, the associated Bianchi identities are derived from the projected classical master equations. Derived brackets induce the associated operations on the twisted Courant algebroids.

Then, we give a representation of T-duality as canonical transformation between QP-manifolds and work it out for the examples of an $S^1$-isometry and three-torus with $H$-flux. Finally, the Poisson Courant algebroid as a model for $R$-flux is reinterpreted in light of our results. We show that it is a special solution of the double field theory section condition realizing $R$-flux on a Poisson manifold.

This paper is organized as follows. In section 2, we give a short review of non-geometric fluxes in string theory. In section 3, an introduction to QP-manifolds, Courant algebroids and double field theory is provided. This will clarify the necessary means to understand the main part, which is section 4. In section 4.1, we derive the fully twisted Courant algebroid with Bianchi identities from canonically transformed Hamiltonians. In section 4.2, the formulation of double field theory via graded symplectic manifolds is introduced. In section 4.3, we discuss the Poisson Courant algebroid with $R$-flux model with respect to double field theory. In section 4.4, we derive the fully twisted double field theory Hamiltonian, that incorporates local expressions for all fluxes and derive Bianchi identities for the winding frame by projection. In section 4.5, a formulation of T-duality in terms of canonical transformations is presented. Section 5 is devoted to discussion of our results and future outlook.

\section{Non-geometric fluxes in string theory}

In this section we give a short introduction to non-geometric backgrounds and their associated non-geometric fluxes.

In general, non-geometric flux backgrounds refer to backgrounds whose mathematical description goes beyond the standard techniques of manifolds. This is in the easiest form observable if one performs T-duality on NS flux backgrounds \cite{Shelton:2005cf,Wecht:2007ngf}. For this let us start with the compactification on a flat six-dimensional torus $T^6$, containing a three-cycle wrapped by the NS three-form $H$-flux. Let the non-zero $H$-flux be denoted by $H_{123}=N$. Then we can take the $B$-field to be $B_{12} = N x^3$. Now, there are several directions to T-dualize. Taking T-duality in the $x^1$-direction leads to a so-called twisted torus background on which there is vanishing $B$-field, and therefore no $H$-flux. One says that the $H$-flux is mapped to the so-called geometric $f$-flux, denoted by $f^1_{23}=N$. The geometric $f$-flux is intimately related to the Scherk-Schwarz fluxes of Scherk-Schwarz compactifications. In the resulting twisted torus background, there are still isometry directions available to T-dualize in. If we T-dualize in the $x^2$-direction, we will be left with what is called a globally non-geometric background. In this case, the $B$-field as well as the metric develop a monodromy, which has to be patched by a T-duality transformation. Therefore, this background is still locally geometric. Such spaces are called T-folds. The associated flux is called $Q$-flux and in this example the resulting background will have $Q^{12}_3=N$, whereas neither NS $H$-flux, nor geometric $f$-flux is present. It turns out that after taking the second T-duality, there is no isometry direction left to T-dualize in. This is in contrast to the fact that we started with background, which possessed this isometry in the beginning. Discussions and an analysis of the resulting structure and backgrounds associated with non-geometric fluxes can be found for example in \cite{Shelton:2005cf,Hull:2004in}. In the literature, this structure was given the name $R$-flux and the associated would-be background is characterized by $R^{123} = N$. It turns out that this background is not even locally geometric, but locally non-geometric, and an analysis via standard manifold and differential geometric methods is impossible.

Let us recall the well-known T-duality chain that has been analyzed in \cite{Shelton:2005cf},
\begin{equation}
	H_{abc} \xleftrightarrow{T_a} f^a_{bc} \xleftrightarrow{T_b} Q^{ab}_c \xleftrightarrow{T_c} R^{abc}, \notag
\end{equation}
where $T_a$ denotes T-duality taken in $x^a$-direction. The authors of \cite{Shelton:2005cf} discussed non-geometric backgrounds of toroidal compactifications of type II string theory from the viewpoint of symmetries of the effective superpotential. More precisely, they considered a six-torus, which factorizes into three identical two-tori divided by a $\mathbb{Z}_2$ and a $\mathbb{Z}_3$ symmetry. This leads to a so-called STU-model, where there are one complex structure modulus $T$, one K\"{a}hler modulus $U$ and one axio-dilaton $S$ for the $T^2$ in the factorization. Imposing T-duality invariance on the resulting $\mathcal{N}=1$ superpotential then leads to the necessity to include further coefficients, that the authors argue to be non-geometric fluxes.

An analysis of the topological aspects of T-duality was carried out in 
\cite{Bouwknegt:2003vb,Bouwknegt:2004tr,Bouwknegt:2010zz}. The action of T-duality on circle bundles was investigated and it turned out that on the topological level T-duality can be seen to exchange the first Chern class of the circle bundle with the background $H$-flux. In this sense, for each circle bundle $E$ with first Chern class $c_1(E)$ and $H$-flux $H$ there exists a T-dual circle bundle $\hat{E}$ with $H$-flux $\hat{H}$, so that $c_1(\hat{E}) = \pi_* H$ and $c_1(E) = \hat{\pi}_* \hat{H}$, where $\pi$ and $\hat{\pi}$ are the respective bundle projections.

A T-duality manifest approach to toroidal string compactifications is given by the so-called double field theory \cite{Hull:2009mi}. The manifest invariance under isometric as well as non-isometric T-duality transformations is established by introducing double coordinates $\tx_i$, which are dual to the standard ones $x^i$. The dual coordinates are interpreted as parameterizing the winding sector of the closed strings wrapping the toroidal cycles. In double field theory, the incorporated fields are simultaneously dependent on the standard and the dual coordinates. The strong point of this theory is that geometric as well as non-geometric backgrounds can be described in a unified manner. However, since the coordinates are doubled, a so-called section condition or strong constraint has to be imposed in order to project down to the physical frame. This constraint is intimately related with the level matching condition, which constraints the modes allowed on a closed string. We will provide a mathematical introduction into double field theory in section 3.2. 

Accounts on the structure of double field theory from the viewpoint of graded symplectic manifolds are given in \cite{Deser:2014mxa,Deser:2014ap}. In \cite{Deser:2014mxa}, a Hamiltonian on a symplectic manifold of degree two was proposed, from which the authors computed the C-bracket using derived brackets. Furthermore, the section condition was deduced from the classical master equation of the Hamiltonian. In \cite{Deser:2014ap}, the author considered $\alpha$'-corrections to the C-bracket. The corrections were mimicked by a Moyal-Weyl star product deformation on the graded manifold at first order. 

\section{Graded symplectic manifolds and double field theory}

In this section, we give an introduction to QP-manifolds and double field theory. First we define the necessary objects and focus on QP-manifolds of degree two and their relation to Courant algebroids. Then we provide a short introduction to the realm of double field theory and the associated differential geometry.

\subsection{QP-manifolds and Courant algebroids}

Fundamental to our analysis is the mathematical structure of a QP-manifold.\footnote{QP-manifold is also called symplectic NQ-manifold.} In this subsection, we provide an introduction to the tools necessary to understand the main text. For details on the background and definitions we refer to \cite{Ikeda:2012laksz}.

A QP-manifold ($\cM$, $\omega$, $Q$) of degree $n$ is a non-negatively graded manifold $\cM$ with a graded symplectic structure $\omega$ of degree $n$ and a homological vector field $Q$ of degree one, such that $L_Q\omega = 0$. A vector field $Q$ is called homological if it is nilpotent, $Q^2 = 0$. In general, $\omega$ is called P-structure and ($\cM$, $\omega$) is the associated P-manifold. The vector field $Q$ is called Q-structure. For a function $f\in \mathcal{C}^{\infty}(M)$, the corresponding Hamiltonian vector field $X_f$ is defined via
\begin{equation}
	\iota_{X_f} = - \delta f,
\end{equation}
where $\delta$ denotes the de Rham differential on $\cM$. The graded symplectic structure $\omega$ defines a graded Poisson bracket via
\begin{equation}
	\{f,g\} \equiv (-1)^{|f|+n+1}\iota_{X_f}\iota_{X_g}\omega,
\end{equation}
where $X_f$ and $X_g$ denote the Hamiltonian vector fields corresponding to $f,g\in\mathcal{C}^{\infty}(\cM)$. 

For any QP-manifold one can find a Hamiltonian function $\Theta\in\mathcal{C}^{\infty}(\cM)$ of degree $n+1$ associated to the homological vector field $Q$ such that
\begin{equation}
	Qf = \{\Theta,f\} \notag
\end{equation}
for $f\in\mathcal{C}^{\infty}(\mathcal{M})$. Then the homological condition on the vector field translates to the so-called classical master equation
\begin{equation}
	Q^2 = 0 \quad \Leftrightarrow \quad \{\Theta,\Theta\} = 0. \notag
\end{equation}
One says that $\Theta$ solves the classical master equation. $\Theta$ is also called homological function or Hamiltonian.

It is well known, that Courant algebroids are in one-to-one correspondence with QP-manifolds of degree two \cite{Roytenberg99}. 

Let us shortly recall the definition of a Courant algebroid and then relate it to QP-manifolds of degree two \cite{lwx, Kosmann-Schwarzbach05}. A Courant algebroid consists of a vector bundle $E$ over a smooth manifold $M$. There are three operations. First, there is a pseudo-Euclidean metric on the fiber, which we denote by $\langle \cdot , \cdot \rangle$. Second, there is a so-called anchor map to the tangent bundle over $M$ given by $\rho:E\rightarrow TM$. Third, there is a so-called Dorfman bracket on the sections of $E$, denoted by $[\cdot , \cdot]_D$. Finally, these operations have to obey the following conditions,
\begin{align}
	[e^1, [e^2, e^3]_D]_D &= [[e^1, e^2]_D, e^3]_D + [e^2, [e^1, e^3]_D]_D, \\
	\rho(e^1)\langle e^2 , e^3 \rangle &= \langle [e^1, e^2]_D, e^3 \rangle + \langle e^2, [e^1, e^3]_D \rangle, \\
	\rho(e^1)\langle e^2 , e^3 \rangle  &= \langle e^1 , [e^2, e^3]_D + [e^3, e^2]_D \rangle,
\end{align}
where $e^1$, $e^2$, $e^3\in \Gamma(E)$.

A general Courant algebroid can be reconstructed from a QP-manifold of degree two as follows. Let $E$ be a vector bundle over a smooth manifold $M$. Consider the graded manifold $\cM = T^*[2]E[1]$. The object $E[1]$ denotes the total space, where the fiber degree is shifted by one.\footnote{The map $[n]$ denotes the shift functor \cite{Ikeda:2012laksz}.} Local coordinates on $\cM$ are ($x^i$, $\eta^a$, $\xi_i$) of degrees ($0$, $1$, $2$). Furthermore, we define an injection of the vector bundle $E$ to $\cM$ via
\begin{align}
	j: TM \oplus E&\rightarrow \cM \notag \\
	j: \left(\frac{\partial}{\partial x^i}, x^i, e^a\right) &\mapsto (\xi_i, x^i, \eta^a). \notag
\end{align}
A general section $e\in\Gamma(E)$ can then be pushed forward via
\begin{equation}
	j_*: e = \alpha_a(x) e^a \mapsto \alpha_a(x) \eta^a,
\end{equation}
where $\alpha_a\in\mathcal{C}^{\infty}(M)$. For the QP-manifold of degree two, the associated graded symplectic structure $\omega$ is of degree two. We assume a fiber metric $\langle \eta^a , \eta^b \rangle = k^{ab}$. Then, the graded symplectic structure is chosen as
\begin{equation}
	\omega = \delta x^i \wedge \delta \xi_i + \frac{1}{2} k_{ab} \, \delta \eta^a \wedge \delta \eta^b.
\end{equation}
We define the Q-structure in terms of the Hamiltonian function via
\begin{equation}
	\Theta = \rho^i_a (x) \xi_i \eta^a + \frac{1}{3!}C_{abc}(x) \eta^a \eta^b \eta^c,
\end{equation}
where $\rho^i_a$, $C_{abc}\in\mathcal{C}^\infty(M)$. In this case, the Hamiltonian function has degree three. If the Hamiltonian function satisfies the classical master equation, $\{ \Theta, \Theta\} = 0$, then ($\cM$, $\omega$, $Q$) defines a QP-manifold of degree two.

In order to reconstruct the Courant algebroid, we have to define the associated three operations. They are defined using the graded Poisson bracket and derived brackets as follows. The pseudo-Euclidean metric is recovered by
\begin{equation}
	\langle e^1 , e^2 \rangle \equiv j^*\{j_* e^1, j_* e^2 \}, \label{CAmetric}
\end{equation}
the Dorfman bracket by
\begin{equation}
	[e^1, e^2]_D \equiv - j^*\{\{ j_* e^1, \Theta\}, j_* e^2 \},
\end{equation}
and the anchor map by
\begin{equation}
	\rho(e) f \equiv - j^* \{\{j_* e, \Theta\}, j_* f \},
\end{equation}
where $e^1$, $e^2$, $e\in\Gamma(E)$ and $f\in\mathcal{C}^\infty(M)$. Due to the classical master equation, the three relations induce a Courant algebroid.

Finally, let us discuss the operation of twisting QP-manifolds. A twist is defined via a canonical transformation. Let ($\mathcal{M}$, $\omega$, $\Theta$) be a QP-manifold of degree $n$ and $\alpha\in\mathcal{C}^{\infty}(\mathcal{M})$ be a smooth function of degree $n$. Then the canonical transformation is defined via exponential adjoint action using the graded Poisson bracket,
\begin{equation}
	e^{\delta_\alpha}f = f + \{f, \alpha\} + \frac{1}{2} \{ \{f, \alpha\}, \alpha\} + \cdots, \notag
\end{equation}
where $f\in\mathcal{C}^{\infty}(\mathcal{M})$ is any smooth function on $\cM$. Since the function $\alpha$ is of the same degree as the graded symplectic structure, the adjoint action is degree-preserving and obeys
\begin{equation}
	\{e^{\delta_\alpha}f, e^{\delta_\alpha}g\} = e^{\delta_\alpha}\{f, g\}, \notag
\end{equation} 
where $f,g\in\mathcal{C}^{\infty}(\cM)$ are smooth functions.

\subsection{Double field theory}

Double field theory \cite{Hull:2009mi} is a manifestly T-duality invariant formulation of the effective theory of the string. See also \cite{Aldazabal:2013dft} for a review on that subject. Double field theory is formulated on spaces obtained after toroidal compactification of string theory. 

The action of T-duality on closed strings compactified on $S^1$ with radius $R$ maps to a dual theory with strings wrapping the dual $\tilde{S}^1$ with radius $\tilde{R} = R^{-1}$. Therefore, we can infer that in the case of a $T^n$-compactification various $S^1$-isometric directions are available to T-dualize in. As we mentioned, investigations of T-duality in the case of $T^6$-compactifications with $H$-flux, lead to the conjecture of non-geometric flux backgrounds \cite{Shelton:2005cf}, which go beyond the standard differential geometry and manifold techniques. The geometry of non-geometric backgrounds \cite{Hull:2004in} plays a fundamental role in double field theory.

Let us consider $D$-dimensional string theory compactified on an $n$-dimensional torus $T^n$. We denote the non-compact coordinates by $x^\mu$, and the compact ones by $x^i$. We also introduce so-called doubled coordinates $\tx_i$, dual to the compact ones. This is the starting point of double field theory, where all fields involved depend on the standard and dual coordinates simultaneously. In the case, where the theory is compactified on a $T^D$, the double coordinates are given by ($\tx_i$, $x^i$) and the T-duality group is $O(D,D)$.

The field content of double field theory is given by the $D$-dimensional metric $g$, the two-form field $B$ and the dilaton field $\phi$. For the discussion in this paper, we will ignore $\phi$. The metric and the $B$-field are rearranged into an $O(D,D)$-tensor via the so-called generalized metric $\mathcal{H}_{MN}$, where $M$ and $N$ run over the entire double space from $1$ to $2D$. The generalized metric parametrized by the geometric subgroup is given by
\begin{equation}
	\mathcal{H}_{MN} = \begin{pmatrix} g^{ij}  & -g^{ik}B_{kj} \\ B_{ik}g^{kj} & g_{ij} - B_{ik}g^{kl} B_{lj} \end{pmatrix}.
\end{equation}

T-duality exchanges momentum modes with winding modes in the T-dualized direction. Since we doubled the coordinates, we also introduce associated momentum modes. We denote the standard momentum mode by $p_i$, whereas we denote its dual momentum mode by $\tp^i$. Throughout the main text, we call the dual momentum mode also winding mode.

Let us consider the generators of the $O(D,D)$-representation. We start with an $O(D,D)$-invariant metric denoted by
\begin{equation}
	\eta_{MN} = \begin{pmatrix} 0 & \delta^i_{\Sp j} \\ \delta^{\Sp j}_i & 0 \end{pmatrix}.
\end{equation}
The invariance property can be stated as
\begin{equation}
	h_M^{\Sp P} \eta_{PQ} h_N^{\Sp Q} = \eta_{MN},
\end{equation}
where $h_M^{\Sp P}$ denote $O(D,D)$-matrices. The generators of $O(D,D)$ are given by diffeomorphisms
\begin{equation}
	h_{M}^{\Sp N} = \begin{pmatrix} E^i_{\Sp j} & 0 \\ 0 & E_i^{\Sp j} \end{pmatrix},
\end{equation}
where $E\in\text{GL}(D)$, $B$-transformations
\begin{equation}
	h_M^{\Sp N} = \begin{pmatrix} \delta^i_{\Sp j} & 0 \\ B_{ij} & \delta_i^{\Sp j} \end{pmatrix}
\end{equation}
and $\beta$-transformations
\begin{equation}
	h_{M}^{\Sp N} = \begin{pmatrix} \delta^i_{\Sp j} & \beta^{ij} \\ 0 & \delta_i^{\Sp j} \end{pmatrix},
\end{equation}
where $B_{ij}$ and $\beta^{ij}$ are antisymmetric tensors. Diffeomorphisms and $B$-transformations generate the so-called geometric subgroup of $O(D,D)$. On this level, all fields depend on both coordinates ($\tx_i$, $x^i$). Then, T-duality acts on the generalized metric $\mathcal{H}_{MN}$, generalized momentum $P^N = (\tp_i, p^i)$ and generalized coordinates $X^N = (\tx_i, x^i)$ via
\begin{align}
	\mathcal{H}_{MN}(X) &\mapsto \mathcal{H}_{PQ}(hX) h_M^{\Sp P} h_N^{\Sp Q}, \\
	P^M &\mapsto h^M_{\Sp N} P^N, \\
	X^M &\mapsto h^M_{\Sp N} X^N,
\end{align}
where $h\in O(D,D)$ and $\tp_i$ denote winding modes, whereas $p^i$ denote momentum modes.

Since from the supergravity point of view, double field theory is an $O(D,D)$-manifestly invariant extension through the introduction of dual variables, there has to be a mechanism to reduce to the physical supergravity frame. This reduction is provided by the so called strong constraint or section condition and is given by
\begin{equation}
	\eta^{MN} \partial_M \partial_N \psi = 0,
\end{equation}
where $\partial_M = (\tpar^i, \partial_i)$ reassembles standard and winding derivatives. $\psi$ denotes any field. This condition is $O(D,D)$-invariant and originates from the level matching condition in string theory \cite{Aldazabal:2013dft}. It can be rewritten in the form, which is useful for the main text of this paper,
\begin{equation}
	\tpar^i \partial_i \psi = 0,
\end{equation}
for any field $\psi$. A reduction of double field theory to the supergravity frame is done by taking all fields to not depend on the winding coordinates, loosely expressed by $\tpar^i = 0$. A reduction to the winding frame is possible by taking $\partial_i = 0$. Mixed reductions are also possible.

Since the metric and $B$-field are rearranged into a generalized object, the respective diffeomorphisms and gauge transformations can be unified using so-called generalized diffeomorphisms with generalized gauge parameter $\xi^M = (\tilde{\lambda}_i , \lambda^i)$. On a generalized vector $V^M$ of weight $\omega(V)$, the action of the generalized Lie derivative along $\xi^M$ is given by
\begin{equation}
	L_\xi V^M = \xi^P \partial_P V^M + (\partial^M \xi_P - \partial_P \xi^M) V^P + \omega(V)\partial_P \xi^P V^M.
\end{equation}
Requiring the closure of gauge transformations we find the relation
\begin{equation}
	[L_{\xi_1}, L_{\xi_2}] = L_{[\xi_1, \xi_2]_C},
\end{equation}
where $[\xi_1, \xi_2]_C$ denotes the so-called C-bracket. It is the antisymmetrization of the so-called D-bracket
\begin{equation}
	[\xi_1, \xi_2]_D = L_{\xi_1}\xi_2.
\end{equation}
Reducing the theory to the supergravity frame using the section condition reduces the C-bracket to the Courant bracket and D-bracket to the Dorfman bracket.

Finally, let us shortly discuss the generalized vielbein formulation of double field theory. We can decompose the generalized metric via
\begin{equation}
	\mathcal{H}_{MN} = E^A_{\Sp M} S_{AB} E^B_{\Sp N},
\end{equation}
where 
\begin{equation}
	S_{AB} = \begin{pmatrix} \eta^{ab} & 0 \\ 0 & \eta_{ab} \end{pmatrix} \notag
\end{equation}
and $E^A_{\Sp M}$ denote generalized vielbeins such that
\begin{equation}
	\eta_{MN} = E^A_{\Sp M} \eta_{AB} E^B_{\Sp N},
\end{equation}
where
\begin{equation}
	\eta_{AB} = \begin{pmatrix} 0 & \delta_{a}^{\Sp b} \\ \delta^{a}_{\Sp b} & 0 \end{pmatrix}. \notag
\end{equation}
The matrix $\eta_{ab}$ denotes the $D$-dimensional metric. $E^A_{\Sp M}$ transforms under generalized diffeomorphisms by
\begin{equation}
	L_\xi E^A_{\Sp M} = \xi^P \partial_P E^A_{\Sp M} + (\partial_M \xi^P - \partial^P \xi_M) E^A_{\Sp P}.
\end{equation}
It can be parametrized by the geometric subgroup of $O(D,D)$ via
\begin{equation}
	E^A_{\Sp M} = \begin{pmatrix} e_a^{\Sp i} & e_a^{\Sp j} B_{ji} \\ 0 & e^a_{\Sp i} \end{pmatrix},
\end{equation}
where $g_{ij} = e^a_{\Sp i} \eta_{ab} e^b_{\Sp j}$ and $e^a_{\Sp j}$ are the vielbeins with respect to the $D$-dimensional metric $g_{ij}$.

If we turn on the $\beta$-field, then we allow for the full set of non-geometric backgrounds and the generalized vielbein is written by \cite{Hassler:2014cc}
\begin{equation}
	E^A_{\Sp M} = \begin{pmatrix} e_a^{\Sp i} & e_a^{\Sp j}B_{ji} \\ e^a_{\Sp j}\beta^{ji} & e^a_{\Sp i} + e^a_{\Sp j}\beta^{jk}B_{ki} \end{pmatrix}.
\end{equation}
In general, backgrounds considered in the framework of the full duality group are globally or even locally non-geometric and contain monodromies patched by full $O(D,D)$-transformations.

\section{Non-geometric fluxes and T-duality via graded symplectic manifolds}

In this section, we present the main results of this paper. First we will derive the fully geometrically and non-geometrically twisted formulation of a Courant algebroid. We derive Bianchi identities and the local expressions of all fluxes by twist of a Hamiltonian. Then we extend our analysis to the setting of double field theory. We discuss the Poisson Courant algebroid as a model for $R$-flux from the perspective of double field theory. Then we derive the Bianchi identities in winding space and the associated local flux expressions. Finally, we define T-duality as canonical transformation acting on the double field theory Hamiltonian. 

\subsection{Courant algebroid with geometric and non-geometric fluxes}

In this section, we derive the Courant algebroid, which incorporates the local expressions of the geometric $H$- and $F$-fluxes as well as non-geometric $Q$- and $R$-fluxes in terms of vielbein $e^{\Sp i}_a$, $B$-field and bivector field $\beta$. We show, that the QP-manifold, which induces this Courant algebroid naturally encodes the Bianchi identities of these fluxes.

Since Courant algebroids are QP-manifolds of degree two, let us start with the simplest QP-manifold and gradually input more ingredients. The QP-manifold, we are considering in this section, is defined on the graded manifold $\mathcal{M}=T^{*}[2]T[1]M$, where $M$ is a smooth manifold. $\mathcal{M}$ is locally parametrized by the coordinates ($x^i$, $q^i$) of degree ($0$, $1$) and their conjugate coordinates ($\xi_i$, $p_i$) of degree ($2$, $1$). The symplectic structure is taken as
\begin{equation}
	\omega = \delta x^i \wedge \delta \xi_i + \delta q^i \wedge \delta p_i.
\end{equation}
It has total degree four. On this QP-manifold, a general Hamiltonian will be of degree three. The simplest non-trivial Hamiltonian, we can write down, is given by
\begin{equation}
	\Theta_{\text{S},0} = q^i \xi_i. \label{S0}
\end{equation}
Let us define an injection of the generalized tangent bundle $TM\oplus T^*M$ over $M$ into $\mathcal{M}$ using the map $j:TM \oplus (TM\oplus T^*M)\rightarrow \mathcal{M}$ via
\begin{align}
	j: \left(\frac{\partial}{\partial x^i}, x^i, dx^i, \partial_i\right) \mapsto (\xi_i, x^i, q^i, p_i). \notag
\end{align}
More precisely, we have the following pullbacks
\begin{align}
	j^*: X^i(x) p_i &\mapsto X^i(x) \partial_i, \notag \\
	j^*: \alpha_i(x) q^i &\mapsto \alpha_i(x) dx^i, \notag
\end{align}
where $X^i, \alpha_i\in\mathcal{C}^{\infty}(M)$. Therefore, a general section $e\in\Gamma(TM\oplus T^*M)$ can be pushed forward to $\mathcal{M}$ by
\begin{equation}
	j_*: e = X^i(x) \partial_i + \alpha_i(x) dx^i \mapsto X^i(x) p_i + \alpha_i(x) q^i. \notag
\end{equation} 

Upon contraction of the Hamiltonian \eqref{S0} with the graded Poisson bracket we can define the de Rham operator $d: \Omega^k(M)\rightarrow\Omega^{k+1}(M)$ by
\begin{equation}
	d\alpha \equiv - j^*\{\Theta_{\text{S},0}, j_*(\alpha)\}
\end{equation}
for any $k$-form $\alpha\in\Omega^k(M)$. The pushforward of a $k$-form $\alpha$ is naturally given by
\begin{equation}
	j_* : \alpha = \frac{1}{k!}\alpha_{i_{1}\cdots i_{k}}(x)dx^{i_{1}}\wedge\cdots\wedge dx^{i_{k}} \mapsto \frac{1}{k!}\alpha_{i_{1}\cdots i_{k}}(x) q^{i_{1}}\cdots q^{i_{k}}.
\end{equation}
The nilpotency of the operator $d$ is guaranteed by the classical master equation, $\{\Theta_{\text{S},0},\Theta_{\text{S},0}\} = 0$. We conclude, that \eqref{S0} induces a de Rham cohomology on the forms over $M$.

The next step is to include geometric as well as non-geometric fluxes $H$, $F$, $Q$ and $R$ in this formulation. This is done by introducing them into the Hamiltonian. It is well known, that the $H$-twisted Hamiltonian
\begin{equation}
	\Theta_{\text{S},H} = \xi_i q^i + \frac{1}{3!} H_{ijk}(x)q^i q^j q^k \label{SH}
\end{equation}
induces the structure of an $H$-twisted Courant algebroid via the classical master equation and derived bracket construction. 

In an analogous manner we can write down the $R$-twist of the Hamiltonian $\Theta_{\text{S},0}$,
\begin{equation}
	\Theta_{\text{S},R} = \xi_i q^i + \frac{1}{3!} R^{ijk}(x) p_i p_j p_k. \label{SR}
\end{equation}
In \cite{Mylonas:2012pg} it is proposed, that the AKSZ action functional induced by the Hamiltonian \eqref{SR} provides a description for a string propagating in $R$-space. In this case, the string is embedded into the cotangent bundle $T^{*}M$ and its phase space is twisted by the $R$-flux. This leads to a non-associative star product on phase space. However, the classical master equation of $\Theta_{\text{S},R}$ 
implies that $R$ is trivial.

The local expressions of the $H$-, $F$-, $Q$- and $R$-fluxes can be introduced by appropriate twist of the Hamiltonian. Since this formulation makes use of a QP-manifold of degree two, the allowed degree-preserving twists also have degree two. The two obvious twists in this setup, we call $B$-twist or $B$-transformation
\begin{equation}
	\exp(\delta_B) \equiv \exp\left(\frac{1}{2}B_{ij}(x)q^i q^j\right) \notag
\end{equation}
and $\beta$-twist or $\beta$-transformation
\begin{equation}
	\exp(\delta_\beta) \equiv \exp\left(\frac{1}{2}\beta^{ij}(x)p_i p_j\right), \notag
\end{equation}
where $B_{ij}, \beta^{ij}\in\mathcal{C}^{\infty}(M)$. 

The $O(D,D)$-covariant metric of the Courant algebroid on the generalized tangent bundle is induced by the graded Poisson bracket \eqref{CAmetric}. The sections of the bundle correspond to degree one functions on $\cM$ via $j$. We conclude, that symplectomorphisms on $\cM$ induced by twists proportional to $q^2$, $p^2$ and $pq$ are in one-to-one correspondence with the generators of $O(D,D)$. Twists that are proportional to $\xi$ induce derivatives and transform \eqref{CAmetric}. Thus, they do not correspond to generators of $O(D,D)$.

In order to include vielbeins, we introduce a frame bundle of $T[1]M \oplus T^*[1]M$.
Then, the vielbein can be introduced using another twist.
Let $(q^a, p_a)$ be local coordinates on 
$V[1] \oplus V^*[1]$ corresponding to a flat frame, where $V= \bR^D$ is a flat vector space 
of dimension $D = {\rm dim} (M)$, whereas
$(q^i, p_i)$ correspond to a general frame on $T[1]M \oplus T^*[1]M$.
Then, we can introduce twists in $(T[1]M\oplus T^*[1]M)\oplus V[1]\oplus V^*[1]$ by
\begin{align}
	\exp(\delta_e) \equiv \exp(e^{\Sp i}_a (x) q^a p_i), \notag \\
	\exp(\delta_{e^{-1}}) \equiv \exp(e_{\Sp i}^a (x) q^i p_a). \notag
\end{align}
This brings us into the position to introduce vielbein components. The resulting injection into the generalized tangent bundle with frame bundle can then be written using local coordinates as
\begin{align}
	\check{j}: TM\oplus (TM\oplus T^* M)\oplus V\oplus V^* &\rightarrow T^*[2]T[1]M\oplus V[1]\oplus V^*[1] \notag \\
	\left(\frac{\partial}{\partial x^i}, x^i, dx^i, \partial_i, u^a, u_a \right) &\mapsto ( \xi_i, x^i, q^i, p_i, q^a, p_a). \notag
\end{align}
Note that $\{q^a, p_b\} = \delta^a_b$ and $\{q^i, p_j\} = \delta^i_j$ and all other combinations vanish.

Before introducing the vielbein components by twist, let us discuss the concept using $B$- and $\beta$-transformations. The classical master equation of $\Theta_{\text{S},H}$ restricts the $H$-flux to be closed, $H\in H^3(M,\mathbb{R})$. On the other hand, we can induce the local expression for the $H$-flux in $\eqref{SH}$ by $B$-twist of \eqref{S0} via
\begin{equation}
	\exp(-\delta_B)\Theta_{\text{S},0} = q^i \xi_i + \frac{1}{2}\partial_i B_{jk} q^i q^j q^k,
\end{equation}
so that $H_{ijk} = 3\partial_{[i}B_{jk]}$ or $H=dB$.

If we twist the standard space Hamiltonian
$\Theta_{\text{S},0}$ by $\beta$-transformation we arrive at
\begin{equation}
	\exp(-\delta_\beta)\Theta_{\text{S},0} = \xi_i q^i - \beta^{mi} \xi_m p_{i} + \frac{1}{2} \partial_i \beta^{jk} q^{i}p_j p_k + \frac{1}{2}\beta^{im} \partial_m \beta^{jk} p_{i}p_{j}p_{k}. \label{ThetaBeta}
\end{equation}
The twist induces an $R$-flux term $R^{ijk} = 3\beta^{[i|m|} \partial_m \beta^{jk]}$ or $R = \frac{1}{2} [\beta,\beta]_S$, where $[-,-]_S$ denotes the Schouten bracket. One observes that if $[\beta,\beta]_S=0$, then 
\begin{equation}
	\Theta_{\beta} \equiv - \beta^{mi}\xi_m p_{i} + \frac{1}{2} \partial_i \beta^{jk} q^{i}p_j p_k
\end{equation}
induces a Poisson cohomology on the space of multivector fields $C^{\infty}(T^*[1]M) \simeq \Gamma(\wedge^{\bullet} TM)$ with Lichnerowicz-Poisson differential $d_\beta \equiv [\beta,\cdot]_S$. Since $\{\xi_i q^i, \Theta_{\beta}\} = 0$, the resulting cohomology associated to the Hamiltonian \eqref{ThetaBeta} for $\beta$ a Poisson bivector is the total cohomology of the de Rham-Poisson double complex with total differential $D = d + d_\beta$ acting on elements $\phi\in\Gamma(\wedge^\bullet TM \oplus \wedge^\bullet T^* M)$. The cohomology can be generalized to the so-called standard cohomology of the Courant algebroid on $C^{\infty}(T^*[2]T[1]M)$ defined by the Hamiltonian $\Theta$ \cite{Roytenberg:2002nu}.

In the case where the $p^3$-coefficient is nonzero, we can describe the breaking of the Poisson condition by a totally antisymmetric trivector $R\in\Gamma(\Lambda^3 TM)$. Then, the resulting structure is a so-called quasi-Poisson structure $\frac{1}{2}[\beta, \beta]_S = R$.\footnote{The terminology of 'quasi-Poisson' in this paper follows the one in \cite{Blumenhagen:2011ph}. The authors of \cite{AKM} give a different definition of a quasi-Poisson structure.} If one defines the Poisson bracket associated to $\beta$ as
\begin{equation}
	\{f,g\}_\beta = \beta^{ij}\partial_i f \partial_j g,
\end{equation}
for $f,g\in\mathcal{C}^{\infty}(M)$, then the quasi-Poisson structure manifests itself by breaking of the Jacobi identity,
\begin{equation}
	\{ \{f,g\}_\beta, h \}_\beta + \{ \{h,f\}_\beta, g \}_\beta  + \{ \{g,h\}_\beta, f \}_\beta  = R^{ijk}\partial_i f \partial_j g \partial_k h.
\end{equation}

In addition to the $R$-flux term we observe that a $Q$-flux term $Q^{hk}_i = \partial_i \beta^{hk}$ has been induced by the $\beta$-twist. In terms of the Poisson bracket this expression can be rewritten as \cite{Blumenhagen:2012Bi}
\begin{equation}
	\{x^i, x^j\}_\beta = \int Q_k^{ij} dx^k.
\end{equation}
Therefore, the $Q$-flux term can be associated to non-commutative structures on the closed string.

With this knowledge we can rewrite \eqref{ThetaBeta} in the form
\begin{equation}
	\Theta_\beta = q^i \xi_i + \beta^{im}p_i \xi_m + \frac{1}{2}Q^{jk}_i q^i p_j p_k + \frac{1}{3!} R^{ijk} p_i p_j p_k,
\label{QRTheta}
\end{equation}
defining $Q^{jk}_i \equiv \partial_i \beta^{jk}$ and $R^{ijk} \equiv 3\beta^{[i|m|}\partial_m\beta^{jk]}$. The $\beta$-transformation induces a differential on the space of polyvectors, in addition to the de Rham differential on forms. Let us denote the former by introducing $e_\sharp^i \equiv \beta^\sharp e^i = \beta^{ij}\partial_j\in T M$, where $e^i\in T^* M$ is the dual of $e_i$, and the latter by $e_i \equiv \partial_i\in TM$, following the notation of \cite{Blumenhagen:2012Bi}. The classical master equation, $\{\Theta_\beta, \Theta_\beta\} = 0$, then implies the Bianchi identities
\begin{align}
	\partial_{[m}Q^{[jk]}_{i]} &= 0, \\
	3\beta^{[i|m|}\partial_m Q_n^{jk]} - \partial_n R^{[ijk]} + 3Q^{[i|m|}_n Q^{jk]}_m &= 0, \\
	\beta^{[i|m|}\partial_m R^{jkl]} - \frac{3}{2} R^{[ij|m|}Q_m^{kl]} &= 0.
\end{align}
These relations coincide with the Jacobi identities for the commutation relations
\begin{align}
	[e_i, e_j] &= 0, \\
[e_i, e_{\sharp}^j] &= Q_{i}^{jm} e_m, \\
[e_{\sharp}^i, e_{\sharp}^j] &= R^{ijm}e_m + Q_m^{ij} e_\sharp^m,
\end{align}
where $[\cdot,\cdot]$ denotes the usual commutator on the tangent bundle.

In fact, the Hamiltonian \eqref{QRTheta} defines the following Courant
 algebroid on $TM \oplus T^*M$.
Let $X+\alpha, Y + \gamma \in \Gamma(TM \oplus T^*M)$ be sections of the generalized tangent bundle.
The anchor map is defined as
\begin{equation}
\rho(X+\alpha)f \equiv - j^*\{\{ j_*(X+\alpha), \Theta_{\beta}\}, j_*(f)\} 
= (X + \beta^{\sharp} (\alpha) ) f.
\end{equation}
The Dorfman bracket is defined by
\begin{align}
[X + \alpha, Y + \gamma]_D 
& \equiv - j^*\{\{ j_*(X+\alpha), \Theta_{\beta}\}, j_*(Y + \gamma)\} 
\nonumber \\ 
&= [X, Y] + L_X \gamma - \iota_Y d \alpha
+ [\alpha, \gamma]_{\beta} + L^{\beta}_{\alpha} Y 
- \iota_{\gamma} d_{\beta} X + \iota_{\alpha} \iota_{\gamma} R.
\end{align}
Here $\beta^{\sharp}(\alpha)
\equiv \beta^{ij} \alpha_i(x) \frac{\partial}{\partial x^j}$ 
for any one-form $\alpha = \alpha_i(x) dx^i$. $L_{\alpha}^\beta$ denotes the Poisson-Lie derivative along a one-form $\alpha$,
\begin{equation}
	L_\alpha^{\beta}Y \equiv d_\beta \iota_\alpha Y + \iota_\alpha d_\beta Y.
\end{equation}
The operator $d_\beta$ denotes the Lichnerowicz-Poisson differential on polyvector fields and $[\alpha, \gamma]_{\beta}$ is the Koszul bracket,
$[\alpha, \gamma]_{\beta} \equiv L_{\beta^\sharp(\alpha)}\gamma - \iota_{\beta^\sharp(\gamma)}d\alpha$.
Since the pseudo-metric is induced directly from the graded Poisson bracket,
\begin{align}
	\langle X + \alpha, Y + \gamma \rangle &\equiv j^*\{j_*(X + \alpha), j_*(Y + \gamma)\} \notag \\
	&= X(\gamma) + Y(\alpha),
\end{align}
it does not change upon the $\beta$-twist.

In order to provide a fully twisted Hamiltonian, that introduces local expressions for all fluxes in terms of all potentials including the vielbeins, we now introduce two index ranges: $i,j,k,\dots$ will denote curved coordinates and $a,b,c,\dots$ will denote flat coordinates. 

Let us get warmed up with a discussion of the $f$-flux in the geometric supergravity frame. In general, the $f$-flux is called geometric flux and is associated with the torsion-less part of the projected spin connection. It is well known, that the T-dual of a three-torus with standard metric and non-zero $H$-flux is a nilmanifold with zero $H$-flux. The dual torus can be formulated as $S^1$-bundle over $S^2$ with non-trivial spin connection. The appropriate Hamiltonian, that models the dual torus nilmanifold, has to incorporate a notion of vielbein $e^{\Sp i}_a$ and $f$-flux with the correct relation
\begin{equation}
	f^a_{bc} = 2 e^{\Sp i}_{[c}\partial_{i}e^a_{\Sp j}e_{b]}^{\Sp j}. \label{FFlux}
\end{equation}
There are two approaches that facilitate such a correspondence. Firstly, the Hamiltonian
\begin{equation}
	\Theta_{\text{S},F} = e_{c}^{\Sp i} \xi_i q^c + \frac{1}{2}f_{ab}^c q^a q^b p_c \label{SF}
\end{equation}
induces the relations
\begin{align}
	e_{[b}^{\Sp j}\partial_j e_{a]}^{\Sp i} &= -\frac{1}{2}e_c^{\Sp i} f^c_{ab}, \\
	e_{[d}^{\Sp j}\partial_j f^c_{ab]} &= f_{[ab}^e f^c_{|e|d]},
\end{align}
where $[\cdots]$ denotes antisymmetrization with the corresponding combinatorial factor. The first equation becomes \eqref{FFlux} upon introduction of an inverse $e_{\Sp i}^d$ of $e_c^{\Sp i}$ such that $e_c^{\Sp i}e_{\Sp i}^d = \delta_c^d$. We conclude, that the local expression of the $f$-flux is induced by the classical master equation of the Hamiltonian \eqref{SF}. This is in contrast to the $H$-flux and $R$-flux case, where we introduced the local expression of the fluxes directly into the Hamiltonian by an appropriate twist.

Let us investigate the possibility of doing so for the $f$-flux using an appropriate representation of the generator of diffeomorphisms inside $O(D,D)$ as canonical transformation acting on our Hamiltonian. For this, we start with the Hamiltonian \eqref{S0} and twist it successively by $\exp(-\delta_e)$, $\exp(\delta_{e^{-1}})$ and again $\exp(-\delta_e)$. We assume, that both vielbeins are inverse to each other and orthogonal with respect to both contractions, $e_a^{\Sp i}e^a_{\Sp j} = \delta^i_j$ and $e_b^{\Sp i}e^a_{\Sp i} = \delta^a_b$. The result is
\begin{equation}
	\exp(-\delta_e)\exp(\delta_{e^{-1}})\exp(-\delta_e)\Theta_{\text{S},0} = e_a^{\Sp m} q^a \xi_m + e_a^{\Sp m}\partial_m e_b^{\Sp i} e_{\Sp j}^b q^j q^a p_i + e_a^{\Sp m} \partial_m e_b^{\Sp i} e_{\Sp i}^c q^a q^b p_c.
\end{equation}
One first recognizes the anchor map, which now is depending on the vielbein itself. The $q^a q^b p_c$-coefficient gives the local expression of the $f$-flux in terms of the vielbeins
\begin{align}
	\frac{1}{2}f^c_{ab} &=  e_{[a}^{\Sp i} \partial_i e_{b]}^{\Sp j} e_{\Sp j}^c \notag \\
	&= e_{[b}^{\Sp i} \partial_i e_{\Sp j}^c e_{a]}^{\Sp j}.
\end{align}
The mixed index coefficient corresponds to a connection on the frame bundle. Since in general we will derive more complicated expressions of $f$-flux in non-geometric frames, we denote the general flux by $F$ and its basic part by
\begin{equation}
	f^c_{ab} \equiv 2e_{[b}^{\Sp i} \partial_i e_{\Sp j}^c e_{a]}^{\Sp j}.
\end{equation}
Introducing a vielbein basis on the tangent bundle $TM$ by $e_a = e_a^{\Sp i}\partial_i$, we can reformulate this expression by a commutator
\begin{equation}
	 [e_a, e_b] = f_{ab}^c e_c.
\end{equation}
We conclude, that the resulting Hamiltonian represents a local formulation of an $f$-flux background.

The Hamiltonian that contains the local expressions for both the $H$-flux and $f$-flux in terms of the potentials $B$ and $e^{\Sp i}_a$ can be derived by twist in the same manner, via
\begin{align}	&\exp(-\delta_e)\exp(\delta_{e^{-1}})\exp(-\delta_e)\exp(-\delta_B)\Theta_{\text{S},0} \notag \\
&\qquad = e_a^{\Sp m} q^a \xi_m + e_a^{\Sp m}\partial_m e_b^{\Sp i} e_{\Sp j}^b q^j q^a p_i + e_a^{\Sp m} \partial_m e_b^{\Sp i} e_{\Sp i}^c q^a q^b p_c + \frac{1}{2}\partial_i B_{jk} e_a^{\Sp i} e_b^{\Sp j} e_c^{\Sp k} q^a q^b q^c.
\end{align}
This leads to the correct local expression of the $H$-flux in flat coordinates in the geometric supergravity frame via $H_{abc} = 3\partial_{i}B_{jk}e_{[a}^{\Sp i}e_b^{\Sp j} e_{c]}^{\Sp k}$. This can be rewritten as covariant derivative $H_{abc} = 3\partial_{[a}B_{bc]} - f^d_{[ab}B_{|d|c]} \equiv 3\nabla_{[a}B_{bc]}$.

Let us in the following consider the full set of fluxes by inducing their local expressions using the three different types of twist: $B$, $\beta$ and vielbein. For this, we start again with \eqref{S0} and twist it by $B$-field and $\beta$-field successively,
\begin{align}
	\exp(-\delta_\beta)\exp(-\delta_B)\Theta_{\text{S},0} &= \exp(-\delta_\beta)(\xi_i q^i + \frac{1}{2}\partial_i B_{jk} q^i q^j q^k) \notag \\
	&= \xi_i q^i - \xi_m\beta^{mi} p_{i} + \frac{1}{2}\partial_n B_{rs} q^n q^r q^s - ( \partial_m B_{ns} + \frac{1}{2}\partial_s B_{mn})\beta^{si}p_{i}q^m q^n \notag \\
	&\qquad + \left[ \frac{1}{2} \partial_i \beta^{hk} - \frac{1}{2} \partial_{i}B_{rs} \beta^{sh}\beta^{rk}  + \partial_r B_{is}\beta^{sh}\beta^{rk}\right]q^{i}p_h p_k \notag \\
	&\qquad + \left[- \frac{1}{2}\partial_l \beta^{ih}\beta^{lk} + \frac{1}{2}\partial_{n}B_{rs}\beta^{si}\beta^{rh}\beta^{nk}\right] p_{i}p_{h}p_{k}. \end{align}
Finally, we introduce the vielbein freedom via
\begin{align}
&\exp(-\delta_e)\exp(\delta_{e^{-1}})\exp(-\delta_e)\exp(-\delta_\beta)\exp(-\delta_B)\Theta_{\text{S},0} \notag \\
&\quad = \xi_i e_b^{\Sp i} q^b - \xi_m \beta^{ml} e^b_{\Sp l} p_b + e_b^{\Sp m}\partial_m e_a^{\Sp j} e^a_{\Sp i} p_j q^i q^b - \beta^{ml}e^b_{\Sp l}\partial_m e_a^{\Sp j} e^a_{\Sp i} p_j q^i p_b \notag \\
&\qquad + e_{c}^{\Sp m} \partial_m e_a^{\Sp j} e^b_{\Sp j} q^a p_b q^c - \beta^{ml} e^c_{\Sp l} \partial_m e_a^{\Sp j} e^b_{\Sp j} q^a p_b p_c \notag \\
&\qquad + \frac{1}{2}\partial_n B_{rs} e^{\Sp n}_a e^{\Sp r}_b e^{\Sp s}_c q^a q^b q^c - ( \partial_m B_{ns} + \frac{1}{2}\partial_s B_{mn})\beta^{si}e_{\Sp i}^a e^{\Sp m}_b e^{\Sp n}_c p_{a}q^b q^c \notag \\
	&\qquad + \left[ \frac{1}{2} \partial_i \beta^{hk} - \frac{1}{2} \partial_{i}B_{rs} \beta^{sh}\beta^{rk}  + \partial_r B_{is}\beta^{sh}\beta^{rk}\right] e^{\Sp i}_a e_{\Sp h}^b e_{\Sp k}^c q^{a}p_b p_c \notag \\
	&\qquad + \left[- \frac{1}{2}\partial_l \beta^{ih}\beta^{lk} + \frac{1}{2}\partial_{n}B_{rs}\beta^{si}\beta^{rh}\beta^{nk}\right] e_{\Sp i}^a e_{\Sp h}^b e_{\Sp k}^c p_{a}p_{b}p_{c} \notag \\
&\quad = e_b^{\Sp i} \xi_i q^b - \beta^{ml} e^b_{\Sp l} \xi_m p_b + \beta^{ml} e^b_{\Sp l} \partial_m e_a^{\Sp j} e^a_{\sp i} q^i p_j p_b + e_b^{\Sp m} \partial_m e_a^{\Sp j} e^a_{\Sp i} q^i q^b p_j \notag \\
&\qquad + \frac{1}{2}\partial_n B_{rs} e^{\Sp n}_a e^{\Sp r}_b e^{\Sp s}_c q^a q^b q^c + \left[e_b^{\Sp m} \partial_m e_c^{\Sp j} e^a_{\Sp j} - ( \partial_m B_{ns} + \frac{1}{2}\partial_s B_{mn})\beta^{si}e_{\Sp i}^a e^{\Sp m}_b e^{\Sp n}_c \right] p_{a}q^b q^c \notag \\
&\qquad \left[- \beta^{ml} e^c_{\Sp l} \partial_m e_a^{\Sp j} e^b_{\Sp j} + \left[ \frac{1}{2} \partial_i \beta^{hk} - \frac{1}{2} \partial_{i}B_{rs} \beta^{sh}\beta^{rk}  + \partial_r B_{is}\beta^{sh}\beta^{rk}\right] e^{\Sp i}_a e_{\Sp h}^b e_{\Sp k}^c \right] q^{a}p_b p_c \notag \\
&\qquad + \left[- \frac{1}{2}\partial_l \beta^{ih}\beta^{lk} + \frac{1}{2}\partial_{n}B_{rs}\beta^{si}\beta^{rh}\beta^{nk}\right] e_{\Sp i}^a e_{\Sp h}^b e_{\Sp k}^c p_{a}p_{b}p_{c}. \label{ThetaBBetaE}
\end{align}
The twisted Hamiltonian incorporates the local expressions for all fluxes in terms of the potentials $B$, $\beta$ and $e^a_{\Sp i}$ in the supergravity frame.

Let us rewrite \eqref{ThetaBBetaE} by
\begin{align}
	\Theta_{B\beta e} &= e_b^{\Sp i} q^b \xi_i + e^b_{\Sp l} \beta^{lm} p_b \xi_m - e^b_{\Sp l} \beta^{lm}\partial_m e_a^{\Sp j} e^a_{\Sp i} q^i p_j p_b + e_b^{\Sp m} \partial_m e^{\Sp j}_a e^a_{\Sp i} q^i q^b p_j \notag \\
	&\qquad + \frac{1}{3!}H_{abc} q^a q^b q^c + \frac{1}{2} F^a_{bc} p_a q^b q^c + \frac{1}{2} Q_a^{bc} q^a p_b p_c + \frac{1}{3!} R^{abc}p_a p_b p_c,
\label{HFQRTheta}
\end{align}
by defining
\begin{align}
	H_{abc} &= 3\nabla_{[a}B_{bc]}, \label{FluxFirst} \\
	F^a_{bc} &= f^a_{bc} - H_{mns}\beta^{si} e^a_{\Sp i} e_{b}^{\Sp m} e_{c}^{\Sp n}, \\
	f^a_{bc} &=  2 e_{[b}^{\Sp m} \partial_m e_{c]}^{\Sp j} e^a_{\Sp j}, \\
	H_{mns} &= 3\partial_{[m}B_{ns]}, \\
	Q_a^{bc} &= \partial_a\beta^{bc} + f_{ad}^b \beta^{dc} - f_{ad}^c \beta^{db} + H_{isr}\beta^{sh}\beta^{rk}e_a^{\Sp i}e^b_{\Sp h} e^c_{\Sp k}, \\
	R^{abc} &= 3(\beta^{[a|m|}\partial_m\beta^{bc]} + f_{mn}^{[a}\beta^{b|m|}\beta^{c]n}) - H_{mns}\beta^{mi}\beta^{nh}\beta^{sk}e^{a}_{\Sp i} e^b_{\Sp h}e^{c}_{\Sp k}. \label{FluxLast}
\end{align}
The classical master equation of this Hamiltonian encodes the Jacobi identities for the commutators
\begin{align}
	[e_a, e_b] &= F_{ab}^c e_c + H_{abc} e^c_\sharp, \\
	[e_a, e_\sharp^b] &= Q_a^{bc} e_c - F_{ac}^b e_\sharp^c, \\
	[e_\sharp^a, e_\sharp^b] &= R^{abc} e_c + Q_c^{ab} e_\sharp^c,
\end{align}
as well as the Bianchi identity for the $H$-flux, where we defined $e_a \equiv e_a^{\Sp i}\partial_i$ and $e_\sharp^a = \beta^\sharp e^a = \beta^{ab} e_b$ with $\beta^{ab} \equiv e^a_{\Sp i} e^b_{\Sp j} \beta^{ij}$ following \cite{Blumenhagen:2012Bi}. This gives the following Bianchi identities,
\begin{align}
	e_{[a}^{\Sp m}\partial_{|m|} H_{bcd]} - \frac{3}{2}F_{[ab}^e H_{|e|cd]} &= 0,  \label{BianchiFirst} \\
	e^{[a}_{\Sp l}\beta^{|lm|}\partial_m R^{bcd]} - \frac{3}{2}Q^{[ab}_e R^{|e|cd]} &= 0, \\
	e^d_{\Sp l} \beta^{ln}\partial_n H_{[abc]} - 3 e_{[a}^{\Sp n}\partial_n F_{bc]}^d - 3H_{e[ab}Q^{ed}_{c]} + 3F^d_{e[a} F^e_{bc]} &= 0, \\
	-2e^{[c}_{\Sp l}\beta^{|ln|}\partial_n F^{d]}_{[ab]} - 2e^{\Sp n}_{[a}\partial_n Q_{b]}^{[cd]} + H_{e[ab]}R^{e[cd]} + Q^{[cd]}_e F^e_{[ab]} + F^{[c}_{e[a}Q^{|e|d]}_{b]} &= 0, \\
	3e^{[b}_{\Sp l}\beta^{|ln|} \partial_n Q^{cd]}_a - e_a^{\Sp n} \partial_n R^{[bcd]} + 3 F_{ea}^{[b} R^{|e|cd]} - 3 Q^{[bc}_e Q^{|e|d]}_a &= 0. \label{BianchiLast}
\end{align}
Let us derive the operations on the Courant algebroid induced by \eqref{HFQRTheta}.
For sections of the generalized frame bundle, $X + \alpha = X^a \partial_a + \alpha_a dx^a$, the anchor map is given by 
\begin{align}
	\rho(X+\alpha)f 
	&\equiv - j^*\{\{ j_*(X+\alpha), \Theta_{B\beta e}\}, j_*(f)\} \notag \\
	&= (X^a e_a^{\Sp m} \partial_m + \alpha_a \beta^{am} \partial_m) f \notag \\
	&= (X + \beta^{\sharp} (\alpha) ) f, \label{NGAnchor}
\end{align}
where $f\in\mathcal{C}^{\infty}(M)$.
For readability, we derive the Dorfman bracket step by step. Let us start with the Dorfman bracket of two vectors $X=X^a\partial_a$ and $Y=Y^b\partial_b$. We get
\begin{align}
	[X, Y]_D &\equiv - j^*\{\{ j_*(X), \Theta_{B\beta e}\}, j_*(Y)\} \notag \\
	&= [X,Y]^\nabla - \beta^\sharp(\iota_Y \iota_X H) + \iota_Y \iota_X H \notag \\
	&= [X,Y]^\nabla_H + \iota_Y \iota_X H,
\end{align}
where we defined $[X,Y]^\nabla$ as the covariant Lie bracket with covariant derivative $\nabla_a X^b = \partial_a X^b + \Gamma^b_{ae}X^e$, where $\Gamma^b_{ae}$ denotes the Weitzenb\"{o}ck connection related to the geometric $f$-flux by $f_{ae}^b = 2\Gamma^b_{[ae]}$. Furthermore, we defined the covariant $H$-twisted Lie bracket by
\begin{equation}
	[X,Y]^\nabla_H \equiv [X,Y]^\nabla - \beta^\sharp(\iota_Y \iota_X H).
\end{equation}
The evaluation of the Dorfman bracket of two forms $\alpha$ and $\gamma$ leads to
\begin{align}
	[\alpha, \gamma]_D &\equiv - j^*\{\{ j_*(\alpha), \Theta_{B\beta e}\}, j_*(\gamma)\} \notag \\
	&= L_{\beta^\sharp(\alpha)}^{\nabla}\gamma -  \iota_{\beta^\sharp(\gamma)}\nabla\alpha + \iota_{\beta^\sharp(\gamma)}\iota_{\beta^\sharp(\alpha)}H + \iota_\gamma \iota_\alpha R \notag \\
	&= [\alpha,\gamma]^\nabla_{\beta,H} + \iota_\gamma \iota_\alpha R,
\end{align}
where we defined the covariant $H$-twisted Koszul bracket by
\begin{equation}
	[\alpha,\gamma]^\nabla_{\beta,H} \equiv [\alpha,\gamma]^\nabla_\beta + \iota_{\beta^\sharp(\gamma)}\iota_{\beta^\sharp(\alpha)}H.
\end{equation}
The covariant Koszul bracket is defined by
\begin{equation}
	[\alpha,\gamma]^\nabla_\beta \equiv L_{\beta^\sharp(\alpha)}^{\nabla}\gamma -  \iota_{\beta^\sharp(\gamma)}\nabla\alpha,
\end{equation}
using the covariant Lie derivative $L^\nabla_X$ along a vector $X$ acting on forms given by
\begin{equation}
	L^\nabla_X = \nabla\iota_X + \iota_X\nabla,
\end{equation}
where $\nabla$ acts on a one-form $\gamma$ by $\nabla\gamma = \partial_a\gamma_b dx^a\wedge dx^b - \Gamma^d_{ab}\gamma_d dx^a \wedge dx^b$. The mixed Dorfman brackets can be evaluated leading to
\begin{align}
	[\alpha,Y]_D &\equiv - j^*\{\{ j_*(\alpha), \Theta_{B\beta e}\}, j_*(Y)\} \notag \\
	&= -\iota_Y\nabla\alpha + \iota_Y\iota_{\beta^\sharp(\alpha)}H + L_\alpha^{\nabla,\beta}Y - \beta^\sharp(\iota_Y\iota_{\beta^\sharp(\alpha)}H)
\end{align}
and
\begin{align}
	[X,\gamma]_D &\equiv - j^*\{\{ j_*(X), \Theta_{B\beta e}\}, j_*(\gamma)\} \notag \\
	&= L_X^\nabla\gamma + \iota_{\beta^\sharp(\gamma)}\iota_X H - \iota_\gamma \nabla_\beta X - \beta^\sharp(\iota_{\beta^\sharp(\gamma)}\iota_X H),
\end{align}
where $L_\alpha^{\nabla,\beta}$ denotes the covariant Poisson-Lie derivative defined by
\begin{equation}
	L_\alpha^{\nabla,\beta} \equiv \nabla_\beta \iota_\alpha + \iota_\alpha \nabla_\beta.
\end{equation}
The symbol $\nabla_\beta$ denotes the covariant Lichnerowicz-Poisson differential. Finally, we summarize the full Dorfman bracket,
\begin{align}
	[X + \alpha,Y + \gamma]_D &= [X,Y]^\nabla_H + [\alpha,\gamma]_{\beta,H}^\nabla - \iota_\gamma\nabla_\beta X - \iota_Y\nabla\alpha + L_X^\nabla \gamma + L^{\nabla,\beta}_\alpha Y + \iota_Y\iota_X H \notag \\
	&\qquad + \iota_Y\iota_{\beta^\sharp(\alpha)}H + \iota_{\beta^\sharp(\gamma)}\iota_X H - \beta^{\sharp}(\iota_Y\iota_{\beta^\sharp(\alpha)}H) - \beta^\sharp(\iota_{\beta^\sharp(\gamma)}\iota_X H) + \iota_\gamma \iota_\alpha R. \label{NGDorfman}
\end{align}
We observe, that a lot of terms involve a $\beta^\sharp$-lift, and conclude, that the existence of the bivector $\beta$ is crucial for the mixed vector-form contracted twists. This concludes the derivation of the fully twisted Courant algebroid in the supergravity frame.

\subsection{Double field theory via graded symplectic manifolds}

In the previous section we showed how to derive the local expressions of the fluxes $H$, $F$, $Q$ and $R$ in terms of their potentials $B$, $\beta$ and $e_{\Sp i}^a$. We discussed the cohomological properties of the associated Hamiltonians as well as twisted Courant algebroid structures in terms of derived brackets. Finally, we derived the Bianchi identities incorporating all fluxes. In this section, we step into the realm of double field theory. For this, the smooth underlying manifold $M$ is doubled by introducing what we want to call winding manifold $\tM$.

Let us go into more detail. The graded manifold associated to the double space is modeled via a P-manifold ($\hcM = T^*[2]T[1]\hM$, $\omega$) of degree two with a vector field $Q$ of degree one and $\hM$ an even-dimensional smooth manifold. For our purpose we take $\hM = M \times \tM$, where $M$ and $\tM$ are smooth manifolds. $M$ will be interpreted as standard spacetime and $\tM$ as its double. We choose $Q$ so that $L_Q \omega = 0$. Then ($\hcM = T^*[2]T[1]\hM$, $\omega$, $Q$) is called a pre-QP-manifold \cite{DeserSaemann:2016}. In other words, a pre-QP-manifold is a QP-manifold, where the nilpotency of the homological vector field is weakened.

Moreover, we assume the existence of an $O(D,D)$-invariant metric $\eta_{MN}$ on the fibers. Then, the manifold $\hcM$ is parametrized by the local coordinates ($x^M = (x^i, \tx_i)$, $q^M = (q^i, \tq_i)$, $p_M = (p_i, \tp^i)$, $\xi_M = (\xi_i, \txi^i)$) of degrees ($0$, $1$, $1$, $2$).

The local expression of the graded symplectic structure $\omega$ is given by
\begin{align}
	\omega &= \delta x^M \wedge \delta \xi_M + \delta q^M \wedge \delta p_M \notag \\
	&= \delta x^i \wedge \delta \xi_i + \delta \tx_i \wedge \delta \txi^i + \delta q^i \wedge \delta p_i + \delta \tq_i \wedge \delta \tp^i.
\end{align}
It is a two-form of degree two.

Also in this case we introduce a map $\hj: T\hM \oplus (T\hM\oplus T^*\hM)  \rightarrow\hcM$ that injects the generalized double tangent bundle to $\hcM$ by
\begin{equation}
	\hj: \left(\frac{\partial}{\partial x^i}, \frac{\partial}{\partial \tx_i}, x^i, \tx_i, dx^i, d\tx_i, \partial_i, \tpar^i \right) \mapsto (\xi_i, \txi^i, x^i, \tx_i, q^i, \tq_i, p_i, \tp^i).
\end{equation}

Finally, the vector field $Q$ defines the Hamiltonian $\Theta$ by $Q(-) = \{\Theta,-\}$. Due to the large number of local coordinates, there exist many possible terms that can be incorporated into $\Theta$. 

In order to describe the section condition in double field theory, let us begin by writing down the non-twisted double field theory Hamiltonian
\begin{align}
	\Theta_{\text{DFT},0} &= \xi_M(q^M + \eta^{MN} p_M) \notag \\
	&= \xi_i(q^i + \tp^i) + \txi^i(p_i + \tq_i). \label{DFT0}
\end{align}
The classical master equation, $\{\Theta_{\text{DFT},0}, \Theta_{\text{DFT},0}\} = 0$, gives us the relation
\begin{equation}
	\xi_i \txi^i = 0. \label{SC}
\end{equation}
This equation can be regarded as the double field theory section condition via \cite{Deser:2014mxa}
\begin{equation}
	\{\{f,\{\Theta_{\text{DFT},0}, \Theta_{\text{DFT},0}\}\},g\} = 0.
\end{equation}
The variables $\xi_i$ and $\txi^i$ induce the derivatives $\partial_i$ and $\tpartial^i$ by the map $\hj$, respectively. In order to solve \eqref{SC}, we choose a proper graded symplectic submanifold of half rank consistent with the $\xi$'s and the injection map, $\hj: \left(dx^i, d\tx_i, \partial_i, \tpar^i \right) \mapsto (q^i, \tq_i, p_i, \tp^i)$.
For example, if we take $\txi^i = 0$, we choose functions $f, g$ dependent only on ($x^i$, $q^i$, $p_i$, $\xi_i$), and therefore a structure subsheaf $C^{\infty}(T^*[2]T[1]M) \subset C^{\infty}(T^*[2]T[1]\hM)$ is selected by setting $\tq_i = \tp^i=0$, where $M$ is the smooth submanifold parameterized by $x^i$.
Here a function of $C^{\infty}(T^*[2]T[1]M)$ is identified as a function on $C^{\infty}(T^*[2]T[1]\hM)$ by the pullback along the natural projection $\proj:\hM \rightarrow M$. $T^*[2]T[1]M$ is a QP-manifold since $Q^2=0$ on $C^{\infty}(T^*[2]T[1]M)$.

\subsection{Poisson Courant algebroid from double space}

In \cite{Bessho:2015tkk} the authors constructed a Courant algebroid on a Poisson manifold serving as a model for $R$-flux. The starting point is the graded manifold $\cM = T^*[2]T[1]M$, where ($M$, $\pi$) is a smooth manifold with Poisson structure $\pi$. In local coordinates on $\cM$, ($x^i$, $q^i$, $p_i$, $\xi_i$) of degree ($0$, $1$, $1$, $2$), the symplectic structure is given as usual by
\begin{equation}
	\omega = \delta x^i \wedge \delta \xi_i + \delta q^i \wedge \delta p_i.
\end{equation}
The Hamiltonian is defined by
\begin{equation}
	\Theta_\pi= \pi^{ij}\xi_i p_j - \frac{1}{2}\frac{\partial \pi^{jk}}{\partial x^i} q^i p_j p_k + \frac{1}{3!}R^{ijk}p_i p_j p_k.
\end{equation}
The classical master equation, $\{\Theta_\pi, \Theta_\pi\} = 0$, induces the relations $[\pi, R]_S = 0$ and $[\pi,\pi]_S = 0$. Therefore, the bivector $\pi$ has to be Poisson and the tri-vector $R$ has to be $d_\pi$-closed, 
where $d_{\pi}$ is the Lichnerowicz-Poisson differential $d_{\pi}(-)=[\pi,-]_S$ acting on the space of multivector fields $\Gamma(\wedge^{\bullet} TM)$. Via the usual injection map $j$ and derived bracket constructions, the Hamiltonian $\Theta_\pi$ induces the Poisson Courant algebroid \cite{Bessho:2015tkk}. 

Obviously, the anchor part of this construction delivers a notion of differential, which is given by the Lichnerowicz-Poisson operator. By direct comparison to the $H$-fluxed Courant algebroid one easily recognizes the following similarities,
\begin{align}
	\pi^{ij}\xi_i p_j - \frac{1}{2}\frac{\partial \pi^{jk}}{\partial x^i} q^i p_j p_k  \quad &\Leftrightarrow \quad \xi_i q^i, \notag \\
	\frac{1}{3!}R^{ijk}p_i p_j p_k \quad &\Leftrightarrow \quad \frac{1}{3!}H_{ijk}q^i q^j q^k, \notag \\
	d_{\pi} R = [\pi, R]_S = 0 \quad &\Leftrightarrow \quad dH = 0, \notag \\
	[\pi,\pi]_S = 0 \quad &\Leftrightarrow \quad d^2 = 0 \notag.
\end{align}
Furthermore, a $B$-twist of the Courant algebroid without $H$-flux is in analogy with a $\beta$-twist of the Poisson Courant algebroid without $R$-flux. The resulting local fluxes are then given by
\begin{equation}
	d_{\pi} R = [\pi, \beta]_S \sim R \quad \Leftrightarrow \quad dB \sim H. \notag
\end{equation}
One concludes that the introduction of Poisson cohomology is necessary to work with a tri-vector flux in the same manner as with a three-form flux using de Rham cohomology. For non-degenerate Poisson structure one finds the well-known map between Poisson and de Rham cohomologies and a relation between both Hamiltonians \cite{Bessho:2015tkk}. 

There are two ways to relate the Poisson Courant algebroid structure to double field theory. The first way is by twist of the double field theory Hamiltonian in the geometric frame. The second way is by direct comparison of the double field theory Hamiltonian with the Hamiltonian realizing the Poisson Courant algebroid.

In order to discuss the former way, let us take local coordinates $(y^i, \tilde{y}_i)$ such that 
$\Theta_{\text{DFT},0} = \eta_i (q^i + \tp^i) + \tilde{\eta}^i (p_i + \tq_i)$,
where $\{y^i, \eta_j \} = \{\tilde{y}_j, \tilde{\eta}^i \} = \delta^i{}_j$.
We can choose a nontrivial physical configuration space of double field theory, a $D$-dimensional submanifold $M \subset \hM$ with local coordinate $x^i$ under the assumption that $M$ has a Poisson structure $\pi$ as follows. Then consider a local coordinate transformation from the double coordinates $(y^i, \tilde{y}_i)$ of $\Theta_{\text{DFT},0}$
to $(x^i, \tx_i)$ with the following Jacobian,
\begin{eqnarray}
\frac{\partial (x, \tx)}{\partial(y, \tilde{y})}
=
\left(
\begin{matrix}
\frac{\partial x^i}{\partial y^j} 
& \frac{\partial x^i}{\partial \tilde{y}_j} 
\\
\frac{\partial \tx_i}{\partial y^j} 
& \frac{\partial \tx_i}{\partial \tilde{y}_j} 
\end{matrix}
\right)
= 
\left(
\begin{matrix}
\delta^i{}_j 
& \pi^{ij}  \\
0 & 
\delta_i{}^j 
\end{matrix}
\right).
\label{diffeofromytox}
\end{eqnarray}
This local coordinate transformation
can be realized as the twist of the original $\Theta_{\text{DFT},0}$
by a canonical function 
$\alpha_{p} = \frac{1}{2} \pi^{ij}(x) p_i p_j$.
The canonical transformation deforms the homological function,
\begin{eqnarray}
\Theta_{\text{DFT},0}^{\prime} 
= e^{\alpha_p} \Theta_{\text{DFT},0}
&=& \xi_i (q^i + \tp^i) + \tilde{\xi}^i (p_i + \tq_i) + \pi^{ij} \xi_i p_j
- \frac{1}{2} \frac{\partial \pi^{jk}}{\partial x^i}(x) (q^i + \tp^i) p_j p_k.
\end{eqnarray}
The section condition is deformed to
\begin{eqnarray}
\tilde{\xi}^i \left(4 \xi_i 
- \frac{1}{2} \frac{\partial \pi^{jk}}{\partial x^i}(x) p_j p_k
\right)
=0.
\label{sectioncondition2}
\end{eqnarray}
The projection to the standard frame recovers the Poisson Courant algebroid with a standard Courant algebroid part without fluxes,
\begin{equation}
	\Theta_{\text{DFT},0}^{\prime}|_{\tx=0} = \Theta_{H=0}
	+ \Theta_{\pi,R=0}.
\end{equation}
Note that, since $\sbv{\Theta_{H=0}}{\Theta_{\pi,R=0}}=0$,
the projected Hamiltonian $\Theta_{\text{DFT},0}^{\prime}|_{\tilde{x}=0}$
defines a double complex.

Now, let us relate the Poisson Courant algebroid by direct comparison of the associated Hamiltonians,
\begin{align}
	\Theta_{\text{DFT},0} &= \xi_i(q^i + \tp^i) + \txi^i (p_i + \tq_i), \notag \\
	\Theta_{\pi,R=0} &= \pi^{ij}(\cx)\cxi_i \cp_j - \frac{1}{2}\frac{\partial \pi^{jk}}{\partial \cx^i}(\cx) \cq^i \cp_j \cp_k, \notag
\end{align}
where we choose a special frame ($\cq_i$, $\cq^i$) for the Poisson Courant algebroid on a $D$-dimensional submanifold $\check{M} \subset \hM$ inside double field theory.
The result is that a Poisson Courant algebroid depending on coordinates $\cx^i$ can be related to double field theory on double coordinates $x^i$ and $\tilde{x}_i$ in the winding frame ($\xi_i = 0$, $q^i = 0$, $p_i = 0$) by the identification
\begin{equation}
	\cq^i = \tp^i, \quad \cp_i = \tq_i, \quad \pi^{ij}(\cx) \cxi_i + \frac{1}{2}\frac{\partial \pi^{jk}}{\partial \cx^i} \cq^i \cp_k  = \txi^j \notag
\end{equation}
and therefore
\begin{equation}
	\tx_i  = \int \pi_{ij}^{-1}(\cx) d\cx^j. \notag
\end{equation}
The term $\frac{1}{2}\frac{\partial \pi^{jk}}{\partial \cx^i} \cq^i \cp_k$ is the Poisson connection induced on the total space. 
Then, the Poisson structure $\pi$ can be seen as a deformation of the double field theory winding frame and the trivector freedom $R$ inside the Poisson Courant algebroid lives inside this deformed winding frame.

\subsection{Introduction of fluxes}

Starting from the double field theory Hamiltonian, the non-twisted supergravity frame Hamiltonian is defined by $\txi^i=0$, whereas
the non-twisted winding frame Hamiltonian by $\xi_i=0$. Then,
$\Theta_{\text{DFT},0}$ reduces to the following Hamiltonians on 
$T^*[2]T[1]M$ and $T^*[2]T[1]\tM$, respectively:
\begin{align}
	\Theta_{\text{S},0} &= q^i \xi_i, \\
	\Theta_{\text{W},0} &= \tq_i \txi^i.
\end{align}
As for the cohomological structure associated with the Hamiltonians, 
$\Theta_{\text{S},0}$ and $\Theta_{\text{W},0}$
induce de Rham cohomologies on the space of forms on standard and winding space, respectively, so that
\begin{align}
	d\alpha &\equiv -\tj^*\{\Theta_{\text{S},0},\tj_*\alpha\}, \\ 
	\tilde{d}\tilde{\alpha} &\equiv -\tj^*\{\Theta_{\text{W},0},\tj_*\tilde{\alpha}\},
\end{align}
for any $k$-forms $\alpha\in\Omega^k(M)$ and $\tilde{\alpha}\in\Omega^k(\tM)$.  

In order to discuss double field theory, which treats all $H$-, $F$-, $Q$- and $R$-fluxes on the same footing, we will derive the fully twisted Hamiltonian incorporating both de Rham differentials $d$ and $\td$. Locally, these fluxes can be written in terms of their potentials: the 2-form $B$-field, the bivector $\beta$-field and vielbeins $e^i_a$.

Then, the $B$- and $\beta$-twisted double field theory Hamiltonian is given by
\begin{align}
	&\exp(-\delta_\beta)\exp(-\delta_B) \Theta_{\text{DFT},0} \notag \\
&\quad = (\xi_i - B_{mi}\txi^m)q^i + (\txi^i - \xi_m\beta^{mi} + \txi^n B_{nm}\beta^{mi})p_{i} \notag \\
	&\qquad + \frac{1}{2}\left[-B_{in}\tpar^i B_{rs} + \partial_n B_{rs}\right] q^n q^r q^s \notag \\
	&\qquad + \left[\frac{1}{2}\tpar^i B_{mn} + (B_{lm}\tpar^l B_{ns} - \partial_m B_{ns} + \frac{1}{2} B_{ls}\tpar^{l}B_{mn} - \frac{1}{2}\partial_s B_{mn})\beta^{si}\right]p_{i}q^m q^n \notag \\
	&\qquad + \left[ \frac{1}{2} \partial_i \beta^{hk} - \frac{1}{2}B_{li}\tpar^l \beta^{hk} + \tpar^h B_{in}\beta^{nk} \right. \notag \\
	&\qquad - \left.  \frac{1}{2}\left[- B_{li}\tpar^l B_{rs} + \partial_{i}B_{rs} - B_{ls}\tpar^l B_{ir} + \partial_s B_{ir} + B_{lr}\tpar^l B_{is} - \partial_r B_{is}\right]\beta^{sh}\beta^{rk}\right]q^{i}p_h p_k \notag \\
	&\qquad + \left[\frac{1}{2}\tpar^i \beta^{hk} - \frac{1}{4}\partial_l \beta^{ih}\beta^{lk} - \frac{1}{4}\beta^{li}\partial_l \beta^{hk} + \frac{1}{4}B_{ln}\tpar^l \beta^{ih}\beta^{nk} \right. \notag \\
	&\qquad + \frac{1}{4}B_{ln}\beta^{ni}\tpar^l \beta^{hk} - \frac{1}{2}\tpar^i B_{mn}\beta^{nh}\beta^{mk} \notag \\
	&\qquad \left. + \frac{1}{3!}(-B_{ln}\tpar^l B_{rs} + \partial_{n}B_{rs} - B_{ls}\tpar^l B_{nr} + \partial_s B_{nr} + B_{lr}\tpar^l B_{ns} - \partial_r B_{ns})\beta^{si}\beta^{rh}\beta^{nk}\right] p_{i}p_{h}p_{k} \notag \\
	&\qquad + \xi_i\tp^i + \txi^i\tq_i + \frac{1}{2}(\partial_i B_{jk} \tp^i + \tpar^i B_{jk} \tq_i)q^j q^k + \frac{1}{2}(\partial_i \beta^{jk} \tp^i + \tpar^i \beta^{jk}\tq_i) p_j p_k \notag \\
	&\qquad - \partial_i B_{jk} \beta^{km} \tp^i q^j p_m - \tpar^i B_{jk} \beta^{km} \tq_i q^j p_m + \frac{1}{2} \partial_i B_{jk} \beta^{jm} \beta^{kn} \tp^i p_m p_n + \frac{1}{2} \tpar^i B_{jk} \beta^{jm} \beta^{kn} \tq_i p_m p_n.	\label{DFT0T}
\end{align}
One recognizes the emergence of the local expressions for all fluxes in terms of their potentials. Furthermore, one recognizes that the anchor map has been twisted,
\begin{equation}
	\Theta_{\text{DFT},\text{A}} = (\xi_i - B_{mi}\txi^m)q^i + (\txi^i - \xi_m\beta^{mi} + \txi^n B_{nm}\beta^{mi})p_{i} + \xi_i \tp^i + \txi^i \tq_i. \label{DFTA}
\end{equation}
The anchor map part inside a Hamiltonian induces the differential on the associated space. In the twisted case, a covariant differential is induced. For example, the expression proportional to $q^i$ inside \eqref{DFTA} induces the covariant differential
\begin{equation}
	D_i = \partial_i + B_{im}\tpar^m \notag
\end{equation}
on $k$-forms over $M$ and the $p_i$-part induces the covariant differential
\begin{equation}
	\tilde{D}^i = \tpar^i - \beta^{mi}\partial_m + B_{nm}\beta^{mi}\tpar^n \notag
\end{equation}
on $k$-forms over $\tM$. Note that we chose a certain order of twists: first $B$-twist, then $\beta$-twist. If we reverse the order of twists, it would change the way of parameterization of the local expressions for the fluxes. Furthermore, we will see that the local expression for the generalized vielbein in double field theory is encoded in the anchor map. Before doing this, we introduce vielbeins in our setup.

We again introduce a frame bundle, in this case for the double space $T[1]\hM \oplus T^*[1]\hM$. Let $(q^a, p_a, \tq_a, \tp^a)$ be local coordinates on $V[1] \oplus V^*[1] \oplus \tilde{V}[1] \oplus \tilde{V}^*[1]$ corresponding to a flat frame, where $\tilde{V} = V = \bR^D$ are flat vector spaces.  $(q^i, p_i, \tq_i, \tp^i)$ correspond to a general frame on $T[1]\hM \oplus T^*[1]\hM$. The injection of the double space frame bundle is then given by
\begin{align}
	\check{\hj}: T\hM \oplus (T\hM \oplus T^*\hM)&\oplus V \oplus V^* \oplus \tilde{V} \oplus \tilde{V}^* \notag \\
	 &\rightarrow T^*[2]T[1]\hM\oplus V[1]\oplus V^*[1] \oplus \tilde{V}[1]\oplus \tilde{V}^*[1] \notag \\
	\left(\frac{\partial}{\partial x^i}, \frac{\partial}{\partial \tx_i}, x^i, \tx_i, dx^i, d\tx_i, \partial_i, \tpar^i, u^a, u_a, \tu_a, \tu^a \right) &\mapsto (\xi_i, \txi^i, x^i, \tx_i, q^i, \tq_i, p_i, \tp^i, q^a, p_a, \tq_a, \tp^a). \notag
\end{align}
Applying the $B$-, $\beta$- and vielbein twists to the untwisted double field theory Hamiltonian \eqref{DFT0} leads to
\begin{align}
&\exp(-\delta_e)\exp(\delta_{e^{-1}})\exp(-\delta_e)\exp(-\delta_\beta)\exp(-\delta_B) \Theta_{\text{DFT},0} \notag \\
&\quad = e_d^{\Sp i} \xi_i q^d - e_d^{\Sp i} B_{mi} \txi^m q^d + e^c_{\Sp l} \txi^l p_c - \beta^{ml} e^c_{\Sp l} \xi_m p_c + e^c_{\Sp l} B_{nm}\beta^{ml}\txi^n p_c \notag \\
	&\qquad  + e_d^{\Sp i}(\partial_i + B_{im} \tpar^m)e_a^{\Sp j} e^a_{\Sp k} p_j q^k q^d + e^c_{\Sp l}(\tpar^l + \beta^{lm}\partial_m + \beta^{lm}B_{mn}\tpar^{n})e_a^{\Sp j} e^a_{\Sp k} p_j q^k p_c \notag \\
	&\qquad + \frac{1}{2}\left[-B_{in}\tpar^i B_{rs} + \partial_n B_{rs}\right] e^{\Sp n}_a e^{\Sp r}_b e^{\Sp s}_c q^a q^b q^c \notag \\
	&\qquad + \left[ e^{\Sp i}_b(\partial_i + B_{im}\tpar^m)e^{\Sp j}_c e^a_{\Sp j} + \frac{1}{2}\tpar^i B_{mn} + \right. \notag \\
	&\qquad \left. + (B_{lm}\tpar^l B_{ns} - \partial_m B_{ns} + \frac{1}{2} B_{ls}\tpar^{l}B_{mn} - \frac{1}{2}\partial_s B_{mn})\beta^{si}\right] e_{\Sp i}^a e^{\Sp m}_b e^{\Sp n}_c p_a q^b q^c \notag \\
	&\qquad + \left[ e^c_{\Sp l}(\tpar^l + \beta^{lm}\partial_m + \beta^{lm} B_{mn}\tpar^n)e_a^{\Sp j} e^b_{\Sp j} + \frac{1}{2} \partial_i \beta^{hk} - \frac{1}{2}B_{li}\tpar^l \beta^{hk} + \tpar^h B_{in}\beta^{nk} \right. \notag \\
	&\qquad - \left.  \frac{1}{2}\left[- B_{li}\tpar^l B_{rs} + \partial_{i}B_{rs} - B_{ls}\tpar^l B_{ir} + \partial_s B_{ir} + B_{lr}\tpar^l B_{is} - \partial_r B_{is}\right]\beta^{sh}\beta^{rk}\right] e^{\Sp i}_a e_{\Sp h}^b e_{\Sp k}^c q^a p_b p_c \notag \\
	&\qquad + \left[\frac{1}{2}\tpar^i \beta^{hk} - \frac{1}{4}\partial_l \beta^{ih}\beta^{lk} - \frac{1}{4}\beta^{li}\partial_l \beta^{hk} + \frac{1}{4}B_{ln}\tpar^l \beta^{ih}\beta^{nk} + \frac{1}{4}B_{ln}\beta^{ni}\tpar^l \beta^{hk} - \frac{1}{2}\tpar^i B_{mn}\beta^{nh}\beta^{mk} \right. \notag \\
	&\qquad \left. + \frac{1}{3!}(-B_{ln}\tpar^l B_{rs} + \partial_{n}B_{rs} - B_{ls}\tpar^l B_{nr} + \partial_s B_{nr} + B_{lr}\tpar^l B_{ns} - \partial_r B_{ns})\beta^{si}\beta^{rh}\beta^{nk}\right] e_{\Sp i}^a e_{\Sp h}^b e_{\Sp k}^c p_{a}p_{b}p_{c} \notag \\
	&\qquad + (\xi_i + \partial_i e_a^{\Sp j} e^a_{\Sp k} p_j q^k + \partial_i e_a^{\Sp j} e^b_{\Sp j} q^a p_b)\tp^i + (\txi^i + \tpar^i e_a^{\Sp j} e^a_{\Sp k} p_j q^k + \tpar^i e_a^{\Sp j}e^b_{\Sp j} q^a p_b)\tq_i \notag \\
	&\qquad + \frac{1}{2}(\partial_i B_{jk} \tp^i + \tpar^i B_{jk} \tq_i)e^{\Sp j}_a e^{\Sp k}_b q^a q^b + \frac{1}{2}(\partial_i \beta^{jk} \tp^i + \tpar^i \beta^{jk}\tq_i) e_{\Sp j}^b e_{\Sp k}^c p_b p_c \notag \\
	&\qquad - \partial_i B_{jk} \beta^{km} e^{\Sp j}_b e_{\Sp m}^c \tp^i q^b p_c - \tpar^i B_{jk} \beta^{km} e^{\Sp j}_b e_{\Sp m}^c \tq_i q^b p_c + \frac{1}{2} \partial_i B_{jk} \beta^{jm} \beta^{kn} e_{\Sp m}^b e_{\Sp n}^c \tp^i p_b p_c \notag \\
	&\qquad + \frac{1}{2} \tpar^i B_{jk} \beta^{jm} \beta^{kn} e_{\Sp m}^b e_{\Sp n}^c \tq_i p_b p_c.\label{DFT0FT}
\end{align}

We can rewrite the expression via
\begin{align}
\tilde{\Theta}_{B\beta e}&= e_d^{\Sp i} \xi_i q^d - e_d^{\Sp i} B_{mi} \txi^m q^d + e^c_{\Sp l} \txi^l p_c - \beta^{ml} e^c_{\Sp l} \xi_m p_c + e^c_{\Sp l} B_{nm}\beta^{ml}\txi^n p_c \notag \\
	&\qquad  + e_d^{\Sp i}(\partial_i + B_{im} \tpar^m)e_a^{\Sp j} e^a_{\Sp k} p_j q^k q^d + e^c_{\Sp l}(\tpar^l + \beta^{lm}\partial_m + \beta^{lm}B_{mn}\tpar^{n})e_a^{\Sp j} e^a_{\Sp k} p_j q^k p_c \notag \\
	&\qquad + (\xi_i + \partial_i e_a^{\Sp j} e^a_{\Sp k} p_j q^k + \partial_i e_a^{\Sp j} e^b_{\Sp j} q^a p_b)\tp^i + (\txi^i + \tpar^i e_a^{\Sp j} e^a_{\Sp k} p_j q^k + \tpar^i e_a^{\Sp j}e^b_{\Sp j} q^a p_b)\tq_i \notag \\
	&\qquad + \frac{1}{2}(\partial_i B_{jk} \tp^i + \tpar^i B_{jk} \tq_i)e^{\Sp j}_a e^{\Sp k}_b q^a q^b + \frac{1}{2}(\partial_i \beta^{jk} \tp^i + \tpar^i \beta^{jk}\tq_i) e_{\Sp j}^b e_{\Sp k}^c p_b p_c \notag \\
	&\qquad - \partial_i B_{jk} \beta^{km} e^{\Sp j}_b e_{\Sp m}^c \tp^i q^b p_c - \tpar^i B_{jk} \beta^{km} e^{\Sp j}_b e_{\Sp m}^c \tq_i q^b p_c + \frac{1}{2} \partial_i B_{jk} \beta^{jm} \beta^{kn} e_{\Sp m}^b e_{\Sp n}^c \tp^i p_b p_c \notag \\
	&\qquad + \frac{1}{2} \tpar^i B_{jk} \beta^{jm} \beta^{kn} e_{\Sp m}^b e_{\Sp n}^c \tq_i p_b p_c \notag \\
	&\qquad + \frac{1}{3!} H_{abc} q^a q^b q^c + \frac{1}{2} F^a_{bc} p_a q^b q^c + \frac{1}{2} Q_a^{bc} q^a p_b p_c + \frac{1}{3!} R^{abc} p_{a}p_{b}p_{c},
\end{align}
by defining
\begin{align}
	H_{abc} &= 3(\nabla_{[a} B_{bc]} + B_{[a|m|}\tpar^m B_{bc]} + \tilde{f}^{mn}_{[a} B_{b|m|} B_{c]n}), \label{First1} \\
	F_{bc}^a &= f^a_{bc} - H_{mns}\beta^{si}e^a_{\Sp i} e_b^{\Sp m} e_c^{\Sp n} + \tpar^a B_{bc} + \tilde{f}^{ad}_b B_{dc} - \tilde{f}_c^{ad}B_{db}, \\
	Q_a^{bc} &= \tilde{f}^{bc}_a +\partial_a\beta^{bc} + f^b_{ad}\beta^{dc} - f^c_{ad}\beta^{db} + H_{isr}\beta^{sh}\beta^{rk}e^{\Sp i}_a e^b_{\Sp h} e^c_{\Sp k} \notag \\
&\qquad + B_{am}\tpar^m \beta^{bc} + \tpar^{[b}B_{ae}\beta^{e|c]} + 2B_{[a|e}\tilde{f}^{be}_{d]}\beta^{dc} - 2B_{[a|e}\tilde{f}^{ce}_{d]}\beta^{db}, \\
	R^{abc} &= 3(\beta^{[a|m|}\partial_m\beta^{bc]} + f^{[a}_{mn}\beta^{b|m|} \beta^{c]n} + \tpar^{[a}\beta^{bc]} - \tilde{f}^{[ab}_d\beta^{|d|c]} \notag \\
	&\qquad  + B_{ln}\tpar^l\beta^{[ab}\beta^{|n|c]} + \tpar^{[a}B_{ed}\beta^{|e|b}\beta^{|d|c]} + \tilde{f}^{[a|e|}_n B_{ed}\beta^{|n|b|}\beta^{|d|c]}) \notag \\
	&\qquad - H_{mns}\beta^{mi}\beta^{nh}\beta^{sk}e^a_{\Sp i}e^b_{\Sp h} e^c_{\Sp k}, \\
	H_{mns} &= 3(\partial_{[m}B_{ns]} + B_{[m|l|}\tpar^l B_{ns]}), \\
	\tilde{f}^{ab}_c &= 2 e^{[a}_{\Sp m} \tpar^m e^{b]}_{\Sp j} e^{\Sp j}_c. \label{Last1}
	\end{align}
The classical master equation then leads to the following relations between the fluxes in the double space
\begin{align}
	e_{[a}^{\Sp i}B_{in}\tpar^n H_{bcd]} + e_{[a}^{\Sp m}\partial_{|m|} H_{bcd]} - \frac{3}{2}F_{[ab}^e H_{|e|cd]} &= 0, \label{First2} \\
	(e^{[a}_{\Sp n} + e^{[a}_{\Sp l}\beta^{lm}B_{mn})\tpar^n R^{bcd]} + e^{[a}_{\Sp l}\beta^{|lm|}\partial_m R^{bcd]} - \frac{3}{2}Q^{[ab}_e R^{|e|cd]} &= 0, \\
	(e^d_{\Sp n} + e^d_{\Sp l}\beta^{lm}B_{mn})\tpar^n H_{[abc]} - 3 e_a^{\Sp i} B_{in}\tpar^n F_{bc}^d + e^d_{\Sp l} \beta^{ln}\partial_n H_{[abc]}& \notag \\
- 3 e_{[a}^{\Sp n}\partial_n F_{bc]}^d - 3H_{e[ab}Q^{ed}_{c]} + 3F^d_{e[a} F^e_{bc]} &= 0, \\
	-2(e_{\Sp n}^{[c} + e^{[c}_{\Sp l}\beta^{lm}B_{mn})\tpar^n F^{d]}_{[ab]} - 2e_{[a}^{\Sp i} B_{in} \tpar^n Q^{[cd]}_{b]} -2e^{[c}_{\Sp l}\beta^{|ln|}\partial_n F_{[ab]}& \notag \\
- 2e^{\Sp n}_{[a}\partial_n Q_{b]}^{[cd]} + H_{e[ab]}R^{e[cd]} + Q^{[cd]}_e F^e_{[ab]} + F^{[c}_{e[a}Q^{|e|d]}_{b]} &= 0, \\
	3(e^{[b}_{\Sp n} + e^{[b}_{\Sp l}\beta^{lm}B_{mn})\tpar^n Q^{cd]}_a - e^{\Sp i}_a B_{in}\tpar^n R^{[bcd]} \qquad \qquad \qquad \qquad& \notag \\
+ 3e^{[b}_{\Sp l}\beta^{ln} \partial_n Q^{cd]} - e_a^{\Sp n} \partial_n R^{[bcd]} + 3 F_{ea}^{[b} R^{|e|cd]} - 3 Q^{[bc}_e Q^{|e|d]}_a &= 0. \label{Last2}
\end{align}
Note that these equations have been derived by twist of the classical master equation
\begin{equation}
	\{\Theta_{\text{DFT},0}, \Theta_{\text{DFT},0}\} \sim \xi^i \txi_i. \notag  
\end{equation}
Since the section condition has not been imposed before twisting, the twisted classical master equation is not solved. In conclusion, the flux expressions \eqref{First1}--\eqref{Last1} do not solve the equations \eqref{First2}--\eqref{Last2}. However, upon solving the section condition, the projected flux expressions solve the projected expressions, which then become Bianchi identities. For instance, by imposing ($\txi^i = 0$, $\tq_i = 0$, $\tp^i = 0$), the set of equations condense to the supergravity frame Bianchi identities presented above. 

On the other hand, solving the section condition by ($\xi_i = 0$, $q^i = 0$, $p_i = 0$) projects to winding frame parametrized flux expressions and their corresponding Bianchi identities. In order to derive local flux expressions in the winding frame, we have to twist by a different set of canonical transformations. We will come to this point below.

The double field theory formulation now brings us into the position to also discuss the winding frame by solving the section condition by ($\xi_i = 0$, $q^i = 0$, $p_i = 0$).
Then, we are left with the untwisted winding frame Hamiltonian
\begin{equation}
	\Theta_{\text{W},0} = \txi^i \tq_i. \label{W0}
\end{equation}
This Hamiltonian induces a de Rham complex in the winding frame. The classical master equation is trivially solved.

From the perspective of the untwisted double field theory Hamiltonian, the standard and winding frames are totally symmetric. The asymmetry between both frames is introduced by the order of twists. Above, we made the choice to first twist by $B$ and then twist by $\beta$. This provided us with a certain parametrization of the fluxes. For instance, the $R$-flux in the supergravity frame (denote $R$-space) has the form of $\frac{1}{2}[\beta,\beta]_S$. In contrast to that, the translation of the $R$-space into a winding frame perspective leads to the Hamiltonian 
\begin{equation}
	\Theta_{\text{W},R} = \txi^i \tq_i + \frac{1}{3!} R^{ijk}(\tx) \tq_i \tq_j \tq_k,
\end{equation}
whose classical master equation in turn forces the $R$-flux to be in third de Rham cohomology over $\hcM$, $R\in H^3(\tM,\mathbb{R})$. This is analogous to having $H\in H^3(M,\mathbb{R})$ due to the classical master equation of \eqref{SH}.

The $H$-twisted Hamiltonian that sees the $H$-flux from the winding frame is given by
\begin{equation}
	\Theta_{\text{W},H} = \txi^i \tq_i + \frac{1}{3!} H_{ijk}(\tx) \tp^i \tp^j \tp^k.
\end{equation}
It is clear that from the winding frame perspective the $H$-flux plays the role of the $R$-flux in standard space, which is resembled by this Hamiltonian.

The winding frame flux expressions are induced from 
\begin{equation}	\exp(-\delta_\te)\exp(\delta_{\te^{-1}})\exp(-\delta_\te)\exp(-\delta_\tbeta)\exp(-\delta_\tB)\Theta_{\text{DFT},0} \notag 
\end{equation}
by solving the section condition via ($\xi_i = 0$, $q^i = 0$, $p_i = 0$),
where
\begin{align}
	\exp(\delta_\tB) &\equiv \exp\left(\frac{1}{2} B_{ij}\tp^i \tp^j\right), \quad \exp(\delta_\te) \equiv \exp(e^{\Sp i}_a \tp^a \tq_i), \notag \\
	\exp(\delta_\tbeta) &\equiv \exp\left(\frac{1}{2} \beta^{ij}\tq_i \tq_j\right), \quad \exp(\delta_{\te^{-1}}) \equiv \exp(e_{\Sp i}^a \tp^i \tq_a). \notag
\end{align}
The result is
\begin{align}
	H_{abc} &= 3(B_{[a|m}\tpar^m B_{bc]} + \tilde{f}^{mn}_{[a} B_{b|m|} B_{c]n}), \label{First3} \\
	F_{bc}^a &= H_{mns}\beta^{si}e^a_{\Sp i} e_b^{\Sp m} e_c^{\Sp n} + \tpar^a B_{bc} + \tilde{f}^{ad}_b B_{dc} - \tilde{f}_c^{ad}B_{db}, \\
	Q_a^{bc} &= \tilde{f}^{bc}_a + H_{isr}\beta^{sh}\beta^{rk}e^{\Sp i}_a e^b_{\Sp h} e^c_{\Sp k} + B_{am}\tpar^m \beta^{bc} + \tpar^{[b}B_{ae}\beta^{e|c]} \notag \\
	&\qquad + 2B_{[a|e}\tilde{f}^{be}_{d]}\beta^{dc} - 2B_{[a|e}\tilde{f}^{ce}_{d]}\beta^{db}, \\
	R^{abc} &= 3(\tpar^{[a}\beta^{bc]} - \tilde{f}^{[ab}_d\beta^{|d|c]} + B_{ln}\tpar^l\beta^{[ab}\beta^{|n|c]} + \tpar^{[a}B_{ed}\beta^{|e|b}\beta^{|d|c]} \notag \\
	&\qquad   + \tilde{f}^{[a|e|}_n B_{ed}\beta^{|n|b|}\beta^{|d|c]}) - H_{mns}\beta^{mi}\beta^{nh}\beta^{sk}e^a_{\Sp i}e^b_{\Sp h} e^c_{\Sp k}, \\
	H_{mns} &= 3B_{[m|l}\tpar^l B_{ns]}, \\
	\tilde{f}^{ab}_c &= 2 e^{[a}_{\Sp m} \tpar^m e^{b]}_{\Sp j} e^{\Sp j}_c. \label{Last3}
	\end{align}
By construction, they obey the following Bianchi identities,
\begin{align}
	e_{[a}^{\Sp i}B_{in}\tpar^n H_{bcd]} - \frac{3}{2}F_{[ab}^e H_{|e|cd]} &= 0, \label{First4} \\
	(e^{[a}_{\Sp n} + e^{[a}_{\Sp l}\beta^{lm}B_{mn})\tpar^n R^{bcd]} - \frac{3}{2}Q^{[ab}_e R^{|e|cd]} &= 0, \\
	(e^d_{\Sp n} + e^d_{\Sp l}\beta^{lm}B_{mn})\tpar^n H_{[abc]} - 3 e_{[a}^{\Sp i} B_{in}\tpar^n F_{bc]}^d - 3H_{e[ab}Q^{ed}_{c]} + 3F^d_{e[a} F^e_{bc]} &= 0, \\
	-2(e_{\Sp n}^{[c} + e^{[c}_{\Sp l}\beta^{lm}B_{mn})\tpar^n F^{d]}_{[ab]} - 2e_{[a}^{\Sp i} B_{in} \tpar^n Q^{[cd]}_{b]} + H_{e[ab]}R^{e[cd]} + Q^{[cd]}_e F^e_{[ab]} + F^{[c}_{e[a}Q^{|e|d]}_{b]} &= 0, \\
	3(e^{[b}_{\Sp n} + e^{[b}_{\Sp l}\beta^{lm}B_{mn})\tpar^n Q^{cd]}_a - e^{\Sp i}_a B_{in}\tpar^n R^{[bcd]} + 3 F_{ea}^{[b} R^{|e|cd]} - 3 Q^{[bc}_e Q^{|e|d]}_a &= 0. \label{Last4}
\end{align}

\subsection{T-duality as canonical transformation}

In order to discuss T-duality, we consider the anchor part of the double field theory Hamiltonian, which is twisted by $B$-, $\beta$ and vielbein fields in the standard as well as in the winding frame,
\begin{align} \exp(-\delta_\te)\exp(\delta_{\te^{-1}})&\exp(-\delta_\te)\exp(-\delta_\tbeta)\exp(-\delta_\tB) \exp(-\delta_e)\exp(\delta_{e^{-1}}) \notag \\
&\times\exp(-\delta_e)\exp(-\delta_\beta)\exp(-\delta_B) \Theta_{\text{DFT},0} = \Theta_{\text{DFT},\text{A}} + \Theta_{\text{DFT},\text{Flux}}. \notag
\end{align}
The flux part $\Theta_{\text{DFT},\text{Flux}}$ is third order in ($q^a$, $p_a$, $\tq_a$, $\tp^a$) and encodes the local expressions of the $H$-, $F$-, $Q$- and $R$-fluxes.
The anchor part $\Theta_{\text{DFT},\text{A}}$ can be rewritten using generalized vielbeins
\begin{align}
	\Theta_{\text{DFT},\text{A}} &= e_a^{\Sp i} \xi_i (q^a +\tp^a) + e_a^{\Sp l} B_{li}\txi^i (q^a + \tp^a) + (e^a_{\Sp i} +  e^a_{\Sp l} B_{im}\beta^{ml}) \txi^i (p_a + \tq_a) + e^a_{\Sp l} \beta^{li} \xi_i (p_a +\tq_a) \notag \\
	&= E_a^{\Sp i} \xi_i (q^a + \tp^a) + E_{ai}\txi^i (q^a + \tp^a) + E^a_{\Sp i} \txi^i (p_a + \tq_a) + E^{ai} \xi_i (p_a + \tq_a),
\end{align}
where we defined
\begin{equation}
	E_a^{\Sp i} \equiv e_a^{\Sp i}, \quad E_{ai} \equiv e_a^{\Sp l} B_{li}, \quad E^a_{\Sp i} \equiv e^a_{\Sp i} +  e^a_{\Sp l} B_{im}\beta^{ml}, \quad E^{ai} \equiv e^a_{\Sp l} \beta^{li}.
\end{equation}
These vielbeins can be reassembled into a generalized vielbein
\begin{equation}
	E^A_{\Sp M} = \begin{pmatrix} E_a^{\Sp i} & E_{ai} \\ E^{ai} & E^a_{\Sp i} \end{pmatrix} = \begin{pmatrix} e_a^{\Sp i} & e_a^{\Sp l} B_{li} \\ e^a_{\Sp l} \beta^{li} & e^a_{\Sp i} +  e^a_{\Sp l} B_{im}\beta^{ml} \end{pmatrix}.
\end{equation}
Introducing the vectors $\Xi^M \equiv (\xi_i, \txi^i)$, $Q_A \equiv (q^a, p_a)$ and $\tilde{P}_A \equiv (\tp^a, \tq_a)$, we can write the anchor part in a manifest form
\begin{equation}
	\Theta_{\text{DFT},\text{A}} = E^A_{\Sp M}(e, B, \beta) \Xi^M (Q_A + \tilde{P}_A).  
\end{equation}
We conclude, that T-duality as $O(D,D)$-transformation relates the generalized vielbeins associated to different backgrounds. A T-duality in $x^k$-direction is the transformation
\begin{equation}
	x^k \leftrightarrow \tx_k, \quad \xi_k \leftrightarrow \txi^k, \quad q^k \leftrightarrow \tq_k, \quad p_k \leftrightarrow \tp^k. \notag
\end{equation}

Let us do some example computations of T-duality on pre-QP-manifolds. The easiest example concerns T-duality on an $S^1$-isometry background without $B$- and $\beta$-fields, where the circle has radius $R$. It is well known, that T-duality maps the radius $R\mapsto R' = \frac{1}{R}$. The corresponding Hamiltonian is given by
\begin{align}
	\Theta_{R} &= e_1^{\Sp 1} \xi_1 (q^1 + \tp^1) + e^{1}_{\Sp 1} \txi^1 (p_1 + \tq_1) \notag \\
	&= R \xi_1 (q^1 + \tp^1) + R^{-1} \txi^1 (p_1 + \tq_1). \label{ThetaR}
\end{align}
We can project into the supergravity frame by taking ($\txi^1 = 0$, $\tq_1 = 0$, $\tp^1 = 0$) leading to
\begin{equation}
	\Theta_{R} = R \xi_1 q^1. \notag
\end{equation}
Applying the transformation described above, the Hamiltonian, which models the T-dual background, is given by
\begin{equation}
	\Theta'_{R^{-1}}	 = R^{-1} \xi_1 (q^1 + \tp^1) + R \txi^1 (p_1 + \tq_1). \label{ThetaRPrime}
\end{equation}
In this case, the projection into the supergravity frame gives
\begin{equation}
	\Theta'_{R^{-1}} = R^{-1} \xi_1 q^1. \notag \label{DualS1}
\end{equation}
We conclude, that this transformation effectively exchanges $R\leftrightarrow R^{-1}$. Alternatively one can derive all T-dual frames from the double field theory Hamiltonian directly by choosing different solutions of the section condition. In the $S^1$-isometry case we could have projected into the winding frame directly by ($\xi_1 = 0$, $q^1 = 0$, $p_1 = 0$) to get the result
\begin{equation}
	\Theta_{R} = R^{-1}\txi^1\tq_1. \notag
\end{equation}
In the next step, the dual variables have to be interpreted as the standard ones leading again to \eqref{DualS1}. This reasoning works in general. If we start with an $H$-flux background, then it looks like $F$-, $Q$- and $R$-flux depending on how we solve the section condition.
 
Finally, let us discuss the case of a three-torus in the directions $i=1,2,3$ with flat metric and $H$-flux $H_{123} = 1$. Its background data is given by
\begin{equation}
	e^{a}_{\Sp i} = \begin{pmatrix} 1 & 0 & 0 \\ 0 & 1 & 0 \\ 0 & 0 & 1 \end{pmatrix}, \quad B_{12} = x^3 = -B_{21}, \quad H_{123} = \partial_{3} B_{12} = 1. \notag
\end{equation}
We can write down the anchor part of this background
\begin{align}
	\Theta_{H,A} &= e_i^{\Sp i} \xi_i (q^i + \tp^i) + e_1^{\Sp 1} B_{12}\txi^2 (q^1 + \tp^1) + e_2^{\Sp 2} B_{21}\txi^1 (q^2 + \tp^2) + e^i_{\Sp i} \txi^i (p_i + \tq_i) \notag \\
	&= \xi_i (q^i + \tp^i) + x^3 \txi^2 (q^1 + \tp^1) - x^3 \txi^1 (q^2 + \tp^2) + \txi^i (p_i + \tq_i).
\end{align}
The flux expressions can be computed by plugging the background local information into the flux part of the Hamiltonian associated to that background geometry. 

We have two isometry directions $x^1$ and $x^2$. Let us T-dualize in $x^1$-direction.  The result is
\begin{equation}
	\Theta_{F,A} = \xi_i (q^i + \tp^i) - x^3 \xi_1 (q^2 + \tp^2) + \txi^i (p_i + \tq_i) + x^3 \txi^2 (p_1 + \tq_1). \label{ThetaF}
\end{equation}
From the coefficients of \eqref{ThetaF} we can read off, which fluxes have been turned on or off:
\begin{equation}
	\check{B} = 0, \quad \check{e}^a_{\Sp i} = \begin{pmatrix} 1 & x^3 & 0 \\ 0 & 1 & 0 \\ 0 & 0 & 1 \end{pmatrix}, \quad \check{g}_{ij} = \begin{pmatrix} 1 & x^3 & 0 \\ x^3 & 1 + (x^3)^2 & 0 \\ 0 & 0 & 1\end{pmatrix}, \quad \check{f}^{1}_{23} = 2\check{e}_{[2}^{\Sp m}\partial_m \check{e}_{c]}^{\Sp j} \check{e}^a_{\Sp j} = 1. \notag
\end{equation}
This background describes a twisted torus. If we take the T-dual of \eqref{ThetaF} into direction of $x^2$ we arrive at
\begin{equation}
	\Theta_{Q,A} = \xi_i (q^i + \tp^i) + \txi^i (p_i + \tq_i) - x^3 \xi_1 (p_2 + \tq_2) + x^3 \xi_2 (p_1 + \tq_1). \label{ThetaQ}
\end{equation}
Again, we can read off from the coefficients of the Hamiltonian the respective local field,
\begin{equation}
	\hat{e}^a_{\Sp i} = \begin{pmatrix} 1 & 0 & 0 \\ 0 & 1 & 0 \\ 0 & 0 & 1 \end{pmatrix}, \quad \hat{\beta}^{12} = x^3, \quad \hat{Q}^{12}_{3} = \partial_3 \hat{\beta}^{12} = 1. \notag
\end{equation}
We conclude, that the second T-dual turned the metric twist into the non-geometric potential $\beta$. Finally, we can take the T-dual in $x^3$-direction leading to
\begin{equation}
	\Theta_{R,A} = \xi_i (q^i + \tp^i) + \txi^i (p_i + \tq_i) - \tx_3 \xi_1 (p_2 + \tq_2) + \tx_3 \xi_2 (p_1 + \tq_1). \label{ThetaR}
\end{equation}
In this case, the former standard coordinate $x^3$ turned into its dual $\tx_3$ and vice versa so that
\begin{equation}
	\bar{e}^a_{\Sp i} = \begin{pmatrix} 1 & 0 & 0 \\ 0 & 1 & 0 \\ 0 & 0 & 1 \end{pmatrix}, \quad	\bar{\beta}^{12} = \tx_3, \quad \bar{R}^{123} = \tpar^3 \bar{\beta}^{12} = 1. \notag
\end{equation}
In general, we can write down any background geometry in terms of the local fields $e$, $B$ and $\beta$ and compute the T-dual background geometry in terms of new fields $e'$, $B'$ and $\beta'$ by using the procedure presented above. We conclude, that we can discuss T-duality by making sole use of canonical transformations on pre-QP-manifolds of degree two. 

T-duality is the change of the solution to the section condition on the pre-QP-manifold $(\wcM, \omega, Q)$. To each solution of the section condition, there is an associated Courant algebroid. T-duality is then a map between different Courant algebroids realized via the respective QP-manifolds associated to different solutions, $T:\cM_1 \rightarrow \cM_2$.

\section{Conclusion and discussion}

In the first part of this paper, we constructed a Hamiltonian on a QP-manifold of degree two, which incorporates the local expressions of NS $H$-flux, $F$-flux and non-geometric $Q$- and $R$-fluxes in terms of vielbeins, $B$-field and $\beta$-bivector. We discussed the cohomological properties of some special cases of this Hamiltonian and deduced the flux Bianchi identities from the classical master equation. Then, we defined the operations on the resulting Courant algebroid using the derived bracket formalism.

In the second part of this paper, we extended our analysis to the double field theory setting by starting from a pre-QP-manifold. Again, by twist of the associated Hamiltonian we could derive all local expressions for the fluxes in the double field theory setting. Through projection to the winding frame we deduced the associated local expressions for all fluxes in winding space and their Bianchi identities. 

We discussed the formerly introduced Hamiltonian of the Poisson Courant algebroid, a Courant algebroid on a Poisson manifold, in light of our results. The Poisson Courant algebroid as a model for trivector $R$-flux turned out to live in the double field theory winding frame deformed by the Poisson structure.

In the third part, we rewrote the anchor part of the Hamiltonian in a T-duality manifest $O(D,D)$-covariant form that resembles the double field theory generalized vielbein. Based on this observation we proposed a representation of T-duality as a canonical transformation between graded symplectic manifolds and computed two simple examples of T-duality in this formulation.

The failure of the double field theory Hamiltonian to obey the classical master equation is measured by the section condition. The conclusion is that the algebra of double field theory does not constitute a Courant algebroid. The projection of the twisted or non-twisted double field theory Hamiltonian onto the standard or winding sector infers the twisted or non-twisted Courant algebroid structure.
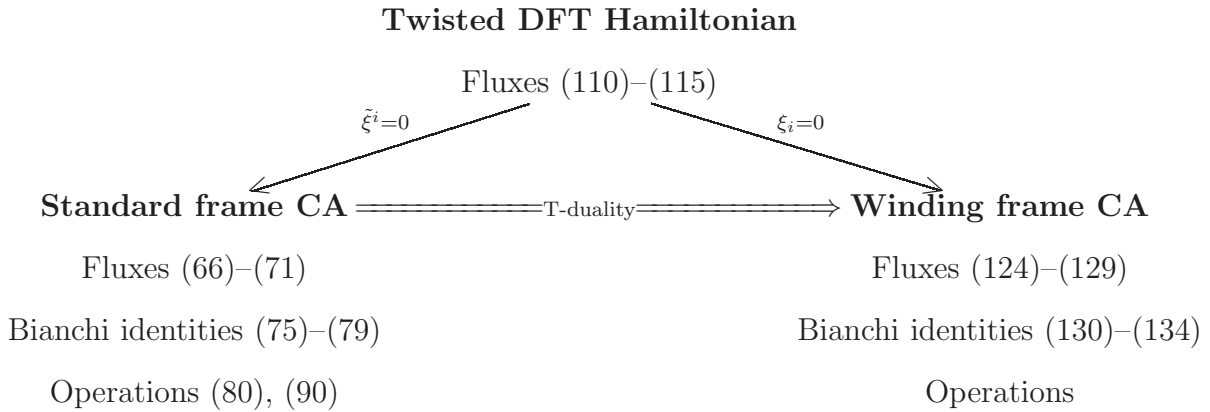
\begin{diagram}
	 &&	\textbf{Twisted DFT Hamiltonian}	& &\\
	&& \text{Fluxes \eqref{First1}--\eqref{Last1}} &&\\
& \ldTo^{\txi^i=0}	& 	&	\rdTo^{\xi_i=0} & \\
\textbf{Standard frame CA} & &  \rImplies~{\text{T-duality}}  & & \textbf{Winding frame CA} \\
\text{Fluxes \eqref{FluxFirst}--\eqref{FluxLast}} & & & & \text{Fluxes \eqref{First3}--\eqref{Last3}}\\
\text{Bianchi identities \eqref{BianchiFirst}--\eqref{BianchiLast}} & & & & \text{Bianchi identities \eqref{First4}--\eqref{Last4}} \\
\text{Operations \eqref{NGAnchor}, \eqref{NGDorfman}}&&&& \text{Operations}
\end{diagram}

The twisted operations of the Courant algebroid in the winding frame remain to be computed by derived brackets. Further topics of future investigation include associated current algebras and topological sigma models. Applications to gravity models can also be thought of. Due to the recent interest in U-duality analogues of double field theory, an extension of this approach to incorporate exceptional duality groups is wished for. Several directions are under exploration. 

\subsection*{Acknowledgments}
\noindent
The authors would like to thank D.~Berman, R.~Blumenhagen, U.~Carow-Watamura, B.~Jur\v{c}o, T.~Kaneko, Y.~Kaneko, O.~Lechtenfeld, C.~S\"amann, P.~Schupp and T.~Strobl
for stimulating discussions and valuable comments.
M.A.~Heller is supported by Japanese Government (MONBUKAGAKUSHO) Scholarship
and
N.~Ikeda is supported by the research promotion program grant 
at Ritsumeikan University.

\end{document}